\documentclass[journal,10pt]{IEEEtran}
\usepackage{fullpage,graphicx,psfrag,amsmath,amsfonts,verbatim,epstopdf}
\usepackage[small,bf]{caption}
\usepackage{color}
\usepackage[utf8]{inputenc}
\usepackage[nolist]{acronym}
\usepackage{cite}

\definecolor{beamer-blue}{RGB}{51,51,178}
\definecolor{beamer-red}{RGB}{189,26,26}

\newtheorem{theorem}{\textbf{Theorem}} 
\newtheorem{lemma}{\textbf{Lemma}} 

\newcounter{assumptionno}
\newenvironment{assumptionno}[1][]{\refstepcounter{assumptionno}\par\medskip
   \textbf{Assumption~\theassumptionno. #1} \rmfamily}{\medskip}

\ifCLASSINFOpdf
\else
\fi
\begin{document}

\begin{acronym}
\acro{a.s.}{\emph{almost surely}}
\acro{EVD}{eigenvalue decomposition}
\acro{i.i.d.}{independent and identically distributed}
\acro{LMS}{least mean square}
\acro{ML}{maximum likelihood}
\acro{MVU}{minimum variance and unbiased}
\acro{SLLN}{strong law of large numbers}
\acro{WSN}{wireless sensor network}
\end{acronym}
%
\title{FADE: Fast and Asymptotically efficient Distributed Estimator for dynamic networks}
%
%
%

\author{Ant\'{o}nio Sim\~{o}es and Jo\~{a}o Xavier,~\IEEEmembership{Member, IEEE}
\thanks{This work was supported in part by the Funda\c{c}\~{a}o para a Ci\^{e}ncia e Tecnologia, Portugal, under Project UID/EEA/50009/2013 and Grant PD/BD/135013/2017. (Corresponding author: Ant\'{o}nio Sim\~{o}es.) The authors are with the Instituto Superior T\'{e}cnico, Universidade de Lisboa, 1649-004 Lisbon, Portugal, and also with the Institute for Systems and Robotics, Laboratory for Robotics and Engineering Systems, 1049-001 Lisbon, Portugal.  (e-mail: {\tt asimoes@isr.ist.utl.pt; jxavier@isr.ist.utl.pt)}.}}

\onecolumn

\maketitle

\begin{abstract}
Consider a set of agents that wish  to estimate a vector of parameters of their mutual interest. For  this estimation goal,  agents can sense and communicate. When sensing, an agent measures (in additive gaussian noise) linear combinations of the unknown vector of parameters. When communicating, an agent can broadcast information to a few other agents, by using  the  channels that happen to be randomly  at its disposal  at the time.

To coordinate the agents towards their estimation goal, we propose a novel algorithm called FADE (Fast and Asymptotically efficient Distributed Estimator), in which agents collaborate at discrete time-steps; at each time-step,  agents sense and communicate just once, while also updating their own estimate of the unknown vector of parameters.

FADE enjoys five attractive features: first, it is an intuitive  estimator, simple to derive; second, it withstands dynamic networks, that is, networks whose communication channels  change randomly over time; third, it is strongly consistent in that, as   time-steps play out, each agent's local estimate converges (almost surely) to the true vector of parameters; fourth, it is both asymptotically unbiased and efficient, which means that, across time,  each agent's estimate becomes unbiased and the mean-square error (MSE) of each agent's estimate vanishes to zero  at the same rate of the MSE of the optimal estimator  at an almighty central node; 
 fifth, and most importantly, when compared with a state-of-art \textit{consensus+innovation} (CI) algorithm, it yields estimates with outstandingly lower mean-square errors, for  the same number of communications---for example, in a sparsely connected network model with $50$ agents, we find through numerical simulations that the reduction can be dramatic, reaching several orders of magnitude.
\end{abstract}

\begin{IEEEkeywords}
Distributed estimation, linear-gaussian models, dynamic networks, consensus+innovations
\end{IEEEkeywords}

%
\IEEEpeerreviewmaketitle

%
%
%
%

 %




\section{Introduction}
\IEEEPARstart{D}{ata} is increasingly collected  by spatially distributed agents, the term \textit{agent} meaning some  physical device that measures data locally, say, a robot. Moreover, not only is the data at these agents being collected  with at an ever-growing rate and precision (as formerly pricy high-quality sensors such as high-resolution cameras have meanwhile became affordable commodities), but the number of agents themselves collecting the data is soaring. Indeed, a present-day wireless sensor network for agriculture precision easily spans  tenths, if not hundreds, of agents, not to mention the blooming vehicular networks or mobile internet-of-things whose size will escalate to even larger scales~\cite{zhao2004wireless}. This steadily increase in both the volume of data and the number of its collectors, however, is at odds with the usual way of extracting information from data in distributed setups: centralized processing. 

\vspace*{2ex}
\noindent\textbf{Centralized processing.}
In centralized processing, the data collected by the agents is usually first routed in raw form (or maybe slightly digested) to a special \textit{central} agent that then performs the bulk of the needed computations on the incoming data to squeeze out the desired information. Centralized processing is poorly suited to the current trend of big data in distributed multi-agent systems, for centralized processing is too fragile and cumbersome. Fragile because, as soon as the central agent breaks down, the whole infrastructure of agents is rendered pointless, incapable of reasoning from the pouring data; cumbersome because, as data falls on the agents at increasing speeds and volumes, the capacity of the physical pathways that convey the data to the central agent must swell in tandem until, of course, this capacity hits a fundamental limit (such as bandwidth of available wireless channels) and the show stops. This explains why centralized processing is gradually eclipsing, giving way to growing research on a different approach for processing data that meshes better with current trends: distributed processing.

\vspace*{2ex}

\noindent\textbf{Distributed processing.} The vision of distributed processing is to untether the data-collecting agents from the central agent and to have the data-collecting agents recreate themselves the  centralized solution, by properly coordinating the agents  in the form of short messages exchanged locally between them. 

Thus, no central agent exists and no particular agent is key; all equally share the computation burden. As a result, distributed processing is more robust. If an agent breaks down, the whole infrastructure does not come to a halt:  just the particular data stream of the faulty agent ceases to inform the solution,  with the remaining agents keeping to collaborate to reason from  their collective data. In sum,  sudden catastrophe in centralized processing (as happening when the central agent collapses) gives place to graceful degradation in distributed processing.

Distributed processing also does away with  the problem of needing larger and larger capacities for  the physical pathways that relay the voluminous collected data to the central agent, for the simple reason that the central agent is crossed out from the picture. Not that any form of communication from agents becomes unnecessary; agents do need to communicate to coordinate. But distributed processing aims at having the agents to exchange short messages only, each message ideally about the size of the information that is desired to squeeze out of the data---a goal which requires modest capacities for the communication channels.

\vspace*{2ex}

\noindent\textbf{Closest related work on distributed estimation.} Research on distributed processing develops vigorously along several exciting threads, among them  distributed optimization, filtering, detection, and estimation. The literature is too vast to discuss at length here, so we point the reader to references~\cite{kar2009distributed,dimakis2010gossip,SocialLearning_2014,FastConv_Nedic:2017,LargeDev_JX2012,SoummyaKar_GLU,SoummyaKar_ExpFam,cattivelli2010diffusion_KF,sayed2013diffusion,chen2012diffusion,mota2012distributed,jakovetic2014fast,shi2015extra}, a  cross-section of representative work.

This paper contributes to the thread of distributed estimation.
 To put our work in perspective, we now single out from this thread some of the closest research. Most of such research assumes a common backdrop: a set of agents is linked by a communication network, with the agents collecting local data about the same vector of parameters; the challenge is to create an algorithm that coordinates the agents by specifying what they should communicate to each other along time over the available communication channels so that the agents arrive at an estimate of the vector of parameters, preferably an estimate as good as a central node would provide for the same time horizon.

Against this shared backdrop, details vary. For example, \cite{Weng2014efficient} considers a linear-gaussian measurement model at each agent, the measurement matrices being time-variant and the observation noise possibly correlated across agents, though the communication network is assumed static, not dynamic (so, communication channels do not change along time). For this setup, the authors work out a distributed estimator, for which they are able to secure a much desired feature: they prove that the proposed estimator is \textit{efficient}, in that its mean-square error (MSE) decays at the same rate of the MSE of the centralized estimator, as times unfolds.

Scenarios in which the vector of parameters changes over time, possibly over dynamic networks, can be  tackled by the successful   suite of distributed algorithms of the \textit{diffusion} type, developed, for instance, in~\cite{lopes2008diffusion,cattivelli2010diffusion,cattivelli2010diffusion_KF,sayed2013diffusion}. Being able to track time-varying parameters, this type of algorithms is fitted to distributed least-mean squares (LMS) or Kalman filtering applications, applications in which these diffusion algorithms can, in fact,  be proved to be stable in the mean-square sense.

We believe that the work that most resembles ours, however,  is the accomplished \textit{consensus+innovations} (CI) algorithm, put forward in~\cite{SoummyaKar_GLU}. Indeed, similar to what we assume in this paper, the authors of~\cite{SoummyaKar_GLU} consider a static vector of parameters, measured at each agent through a noise-additive linear model, and a dynamic communication network linking the agents that changes randomly over time. Thus, although the CI algorithm is applicable to a broader range of scenarios (e.g., gaussian noise is not assumed in~\cite{SoummyaKar_GLU}), we will use it as a benchmark throughout the paper, both in section~\ref{derivingfade}, where we contrast its form with our FADE algorithm, and in section~\ref{sc:simulation}, where we compare their performance in practice.  
(Other distributed algorithms built around the consensus+innovations principle, and not necessarily for estimation purposes,  can be found in\cite{SoummyaKar_GLU,SoummyaKar_ExpFam,LargeDev_JX2012,kar2013consensus,kar2014distributed}.)

\vspace*{2ex}

\noindent \textbf{Contributions.} We contribute a distributed estimator called FADE (Fast and Asymptotically efficient Distributed Estimator), which coordinates agents  measuring a common vector of parameters in a linear-gaussian model and  communicating with each other over a set of channels that changes randomly over time. FADE is endowed with several assets, both on the theoretical side and on the practical side.

On the theoretical side, FADE is simple to derive, emerging from the central estimator after a couple of intuitive steps. (By comparison, the blueprint of the CI algorithm~\cite{SoummyaKar_GLU}, say, is somehow more involved to arrive at, resorting to the machinery of stochastic approximation with mixed time-scales---perhaps a complexity price to pay for its more general applicability.) More importantly, FADE comes with strong convergence guarantees: it is strongly convergent (estimates converge to the true parameters with probability one); it is asymptotically unbiased (bias of the estimates vanishes with time); and it is asymptotically efficient (MSE of the estimates goes to zero at the same rate of the MSE of the centralized estimator, as time proceeds). Our theoretical proofs use tools from martingale probability theory.

On the practical side, FADE is shown, through numerical simulations, to supply estimates to the agents that are notably more accurate than the ones yielded by the CI algorithm, for the same time period. Said in other words: although both FADE and the CI algorithm are asymptotically efficient, numerical simulations show that FADE reaches the asymptotic optimal performance significantly sooner. As an illustration, in one of the simulations we find that the MSE of FADE is lower than the MSE of CI by five orders of magnitude. 

\vspace*{2ex}

\noindent\textbf{Paper organization.} We organize this paper as follows. In section \ref{sc:state_of_art}, we detail the measurement and communication models, along with usual blanket assumptions, thus stating precisely the problem at hand. The problem is tackled in section~\ref{derivingfade}, in which we derive our solution---the FADE algorithm---whose strong theoretical convergence properties are laid out in section~\ref{tfade}. In section~\ref{sc:simulation}, we compare by numerical simulation the accuracy of FADE with the accuracy of the CI algorithm, both in a dense and a sparse network model.
Section~\ref{sc:conclusion} closes the paper, with parting conclusions. Appendices  give the proofs of theorems stated throughout the paper.

\section{Problem statement} \label{sc:state_of_art}

The vector of parameters that the $N$ agents wish to estimate is $\theta \in {\mathbf R}^d$.


\vspace*{2ex}

\noindent\textbf{The measurement model: how agents measure~$\theta$.}
Each agent measures a linear map of $\theta$ in additive gaussian noise. 
Specifically, at time-step $t = 1, 2, \ldots$, agent $n$ measures 
\begin{equation}
y_n(t) = H_n \theta + v_n(t), \label{eq:obs}
\end{equation}
where  $H_n \in {\mathbf R}^{d_n\times d}$ is the linear map of agent~$n$, and $v_n(t)$ is  gaussian noise, detailed in the following assumption.

\begin{assumptionno}[(Gaussian noise)]
For each agent $n$ and time-step $t$, the random variable $v_n(t)$ is a sample of a gaussian distribution with zero mean and unit covariance, written  $v_n(t) \sim \mathcal{N}(0,I_{d_n})$. These random variables are independent across agents and time; that is, $v_n(t)$ is independent of $v_m(s)$ if $n \neq m$ or $t \neq s$.  \label{assumpnoise}
\end{assumptionno}

As an aside, note that if the covariance of the noise was other than the identity matrix, say, $\Sigma_n \neq I_{d_n}$, then the unit-covariance feature could be restored at once by premultiplying $y_n(t)$ with $\Sigma_n^{-1/2}$  (redefining $H_n$ also in the process). In sum, this convenient assumption entails no loss of generality.

We also need a basic observability assumption that states $\theta$ can be identified from all the agents' measurements. Specifically, stack the observations~\eqref{eq:obs} in the column vector $y(t) = \left( y_1(t)^T, y_2(t)^T, \ldots, y_N(t)^T \right)^T \in {\mathbf R}^{d_1 + d_2 + \cdots + d_N}$; this gives $y(t) = H \theta + v(t)$,
where $H = \begin{bmatrix} H_1^T & H_2^T & \cdots & H_N^T \end{bmatrix}^T \in {\mathbf R}^{(d_1 + d_2 + \cdots + d_N) \times d}$ and $v(t) = \left( v_1(t)^T, v_2(t)^T, \ldots, v_N(t)^T \right)^T$. The vector $y(t)$ can be seen as a network-wide measurement: it collects the measurements at time~$t$ from all the agents.
We assume that $\theta$ is identifiable from these network-wide measurements, which is equivalent to assuming that the network-wide sensing matrix $H$ is full column-rank.

\begin{assumptionno}[(Global observability)]
The matrix $H$ is full column-rank. \label{assumpgo}
\end{assumptionno}

\vspace*{-1ex}
For the scalar case $d {=} d_1 {=} d_2 {=} \cdots {=} 1$, this assumption just means that some sensing gain $H_n \in {\mathbf R}$ is nonzero. 
In general, note that  this assumption exempts the local sensing matrices $H_1, \ldots, H_n$ from any particular structure; e.g., all sensing matrices could even lack full column-rank, which would make $\theta$ unindentifiable from any agent (if $H_n$ is not full column-rank, then agent~$n$ can not discern between $\theta$ from $\theta {+} \delta$, where $\delta$ is any nonzero vector in the kernel of $H_n$). Assumption~\ref{assumpgo} ensures that $\theta$ can be identified whenever agents work as a team.

\vspace*{2ex}
\noindent\textbf{The communication model: how agents can exchange information.} The communication network linking the agents is modelled as an undirected graph that changes randomly over time: at time-step $t$, we call this graph \begin{equation} \mathcal G(t) = \left(\mathcal V, \mathcal E(t)\right). \label{grapht} \end{equation} Here, ${\mathcal V} = \{ 1, 2, \ldots, N \}$ is the set of $N$ agents, and ${\mathcal E}(t)$ is the set of edges available at time step~$t$. An edge between agents $n$ and $m$ models a communication channel between them; note that, because the graph is undirected, the channels are bidirectional (which means that if, at a given time step~$t$, agent $n$ can send information to agent $m$, then the reverse also holds: agent $m$ can send information to agent~$n$ as well).

We let channels appear and disappear randomly over time, to model  agents that move around or data packets that get lost; therefore, the terms of the sequence of edge-sets ${\mathcal E}(1), {\mathcal E}(2), {\mathcal E}(3), \ldots$ from~\eqref{grapht} vary along time. Note, however, that each term in this sequence necessarily takes values in a finite collection of edge-sets, say, a collection with $K$ edge-sets: ${\mathcal E}(t) \in \left\{ {\mathcal E}_1, \ldots, {\mathcal E}_K \right\}$. Such collection is finite because each ${\mathcal E}(t)$ must be an  edge-set over the node-set ${\mathcal V}$ and, since ${\mathcal V}$ is fixed, the  collection of all  edge-sets on ${\mathcal V}$ is itself finite. Of course, for a particular application scenario, not all possible edge-sets need to turn out: in fact, the collection $\left\{ {\mathcal E}_1, \ldots, {\mathcal E}_K \right\}$ is the subset of those edge-sets that have a strictly positive probability of turning out for this application scenario.

Finally, we assume that the sequence ${\mathcal E}(1), {\mathcal E}(2), {\mathcal E}(3), \ldots$ satisfies a standard and basic property called average connectivity. 

\begin{assumptionno}[(Average connectivity)] The random sequence of edge-sets ${\mathcal E}(1), {\mathcal E}(2), {\mathcal E}(3), \ldots$, is independent and identically distributed (i.i.d.). Each edge-set ${\mathcal E}(t)$ takes values in the finite collection $\{ {\mathcal E}_1, \ldots, {\mathcal E}_K \}$ with probability $\pi_k = {\mathbf P}\left( {\mathcal E}(t) = {\mathcal E}_k \right)$, where $\pi_k > 0$ and $\sum_{k = 1}^K \pi_k = 1$.
Moreover, the average edge-set is connected, that is, 
the graph $\left( {\mathcal V}, \bigcup_{k = 1}^K {\mathcal E}_K \right)$ is connected. \label{assump2}
\end{assumptionno}

Recall that a graph is connected if there is a path between any two nodes, the path  consisting possibly of many edges. Assumption~\ref{assump2} means that the \textit{average} communication graph is connected. In other words, the assumption means that if we overlay all edge-sets with positive probability of occurring,  a connected graph results. Of course,  at each time step~$t$ the graph ${\mathcal G}(t) = ({\mathcal V}, {\mathcal E}(t) )$ is allowed to be disconnected; such is the case,  for instance, in single-neighbor gossip-like protocols where only two neighbour agents talk at a time (in such a case, each ${\mathcal E}(t)$ contains only one edge).

\vspace*{2ex}
\noindent\textbf{The problem addressed in this paper.} We address the problem of creating an algorithm that runs at each agent and that, by using the locally available sensing and communication resources, provides an estimate---as accurate as possible---of $\theta$ at each agent. For this problem we propose FADE, a Fast and Asymptotically efficient Distributed Estimator, which we derive in the next section.

\section{Deriving FADE}
\label{derivingfade}

FADE is an intuitive algorithm, simple to derive. For clarity, we first focus in section~\ref{sec:scalar} on a scalar parameter, $\theta \in {\mathbf R}$; the extension to the vector case, $\theta \in {\mathbf R}^d$, is plain and appears in section~\ref{sec:vector}.

\subsection{FADE algorithm for scalar parameter}
\label{sec:scalar}

We derive FADE in a progression of three easy steps, starting from the stance of an almighty central node. 

\vspace*{2ex}
\noindent\textbf{Step 1. The optimal algorithm at a central node.} We start by deriving the optimal estimator. This is the estimator that could run only at a central node---a fictitious, almighty node that would know \textit{instantaneously} the measurements of \textit{all} the agents. 

Such optimal estimator is the minimum variance and unbiased (MVU) estimator or, in our linear-gaussian setup, also the maximum likelihood (ML) estimator. It is given by a weighted combination of the average measurements of the agents; specifically, at time~$t = 1, 2, 3, \ldots$, it is given  by \begin{equation} \widehat \theta(t) = \sum_{
m = 1}^N \frac{1}{N} c_m \overline y_m(t) \label{optimalestimator}, \end{equation}
where $c_m =  \left( (1/N) \sum_{i = 1}^N h_i^2  \right)^{-1} h_m$ is the weight of agent~$m$, and \begin{equation} \overline y_m(t) = \frac{1}{t} \sum_{s = 1}^t y_m(s) \label{avey} \end{equation} is the average measurement of agent~$m$ by time~$t$. We skip the details on how to obtain the optimal estimator~\eqref{optimalestimator} because they are well-known (e.g., see~\cite{kay_estimation_theory}).

Now,  we can write~\eqref{avey} in the recursive form $\overline y_m(t) = \overline y_m(t-1) + ( 1/ t )( y_m(t) - \overline y_m(t-1) )$; and plugging this recursion in~\eqref{optimalestimator}  gives the following update for the optimal estimator:
\begin{equation}
\widehat \theta(t) = \frac{1}{N} \sum_{m = 1}^N \left( \widehat \theta( t - 1 ) + \frac{1}{t}   c_m \left( y_m(t) - \overline y_m(t-1) \right) \right),
\label{optimalestimatorrec}
\end{equation}
for $t = 1, 2, 3, \ldots$, where we set $\widehat \theta ( 0 ) = 0$ and $\overline y_m( 0 ) = 0$ for all~$m$.

Update \eqref{optimalestimatorrec} reveals an interesting feature: at time step~$t$,  the optimal estimator in~\eqref{optimalestimatorrec} needs only to know from each agent~$m$ the number $c_m ( y_m(t) -  \overline y_m(t-1) )$. This means that, besides a central node, the optimal estimator can also run in a certain distributed scenario----a scenario in which the communication graph is static and complete, as we show in the next step.

\vspace*{2ex}
\noindent\textbf{Step 2. The optimal algorithm in a static, complete graph} Consider a static, complete graph. Being static, its edges are fixed over time. Being complete, it contains all possible edges, that is, any pair of agents is linked by an edge. In this graph any given agent $n$ can obtain whatever information it needs from any other agent~$m$ at any time step;  agent~$m$ has only to send that information through the channel linking agents $n$ and $m$. In particular, agent~$m$ can send  $c_m ( y_m(t) - \overline y_m(t-1) )$. So, any agent~$n$ can carry out the optimal update~\eqref{optimalestimatorrec}. 

Letting, then, $\widehat \theta_n(t)$ be the estimate at agent~$n$ in such a graph, we have
\begin{equation}
\widehat \theta_n(t) {=} \frac{1}{N} \sum_{m = 1}^N \left( \widehat \theta_n( t {-} 1 ) {+} \frac{1}{t}   c_m \left( y_m(t) {-} \overline y_m(t{-}1) \right) \right),
\label{agentm}
\end{equation}
with $\widehat \theta_n( 0 ) = 0$. Note that  the agents' estimates keep the same through time, $\widehat \theta_1(t) =  \widehat \theta_2(t) = \cdots = \widehat \theta_N(t)$, and recreate the optimal update in~\eqref{optimalestimatorrec}.

Let us pass to a more convenient vector form. 
Stack all agent's estimates in  the vector $\widehat \theta(t) = \left( \widehat \theta_1(t), \ldots, \widehat \theta_N(t) \right) \in {\mathbf R}^N$. It follows from~\eqref{agentm} that
\begin{equation}
\widehat \theta(t) = J \left( \widehat \theta( t - 1 ) + \frac{1}{t}   C \left( y(t) - \overline y(t) \right) \right),
\label{vecform}
\end{equation}
  where  $J = {\mathbf 1} {\mathbf 1}^T / N$ (with ${\mathbf 1} = \left( 1, 1, \ldots, 1 \right) \in {\mathbf R}^N$) is  the \textit{consensus matrix}, 
$C \in {\mathbf R}^{N\times N}$ is a diagonal matrix with $n$th diagonal entry equal to $c_n$, $y(t) = \left( y_1(t), \ldots, y_N(t) \right) \in {\mathbf R}^N$, and $\overline y(t) = \left( \overline y_1(t), \ldots, \overline y_N(t) \right) \in {\mathbf R}^N$.

\vspace*{2ex}

\noindent\textbf{Step 3. The FADE algorithm in a general graph.}	The optimal estimator~\eqref{vecform} is unable to run in a general graph changing over time. Indeed, let ${\mathcal E}(t)$ be the set of available edges for communication at time step~$t$. At this time step,  only a  few communication channels  typically link a given  agent~$n$ to other agents. Specifically,  agent~$n$ can receive information only from those agents $m$  for which the (undirected) edge $\{  n, m \}$ is in ${\mathcal E}(t)$, a subset  called the \textit{neighborhood} of agent~$n$ at time step~$t$ and  denoted by ${\mathcal N}_n(t) = \left\{ m \neq n \,:\, \{ n, m \} \in {\mathcal E}(t) \right\}$. 
Sadly, the update~\eqref{vecform} requires agent~$n$ to receive information, not just from \textit{some} neighbor agents, but from \textit{all}  agents. As such, the update~\eqref{vecform} cannot run in a general graph.

Inspired by the form of the recursion~\eqref{vecform}, however, we now suggest a simple modification that can run in general graphs. The key idea is to note that the obstruction is just the consensus matrix $J$. Indeed, each entry $(n, m)$ of $J$ is non-zero (to be more specific, $J_{n m} = 1/N$): this makes the update at agent $n$ to depend on the information at any other agent $m$. But, would that entry be zero, and the update at agent~$n$ would no longer depend on the information at agent $m$. 

Our idea now almost suggests itself: simply replace the consensus matrix $J$ in~\eqref{vecform} with a matrix, say $W(t)$, that has the right sparsity at time~$t$. That is, a matrix $W(t)$ where each entry $W_{n m}(t)$ nonzero if and only if agents $n$ and $m$ are neighbors at time~$t$: $W_{n m}(t) \neq 0$ if and only if $\{n, m \} \in {\mathcal E}(t)$. 
We finally arrive at our FADE algorithm:
\begin{equation}
\widehat \theta(t) = W(t) \left( \widehat \theta( t - 1 ) + \frac{1}{t}  C \left( y(t) - \overline y(t-1) \right) \right),
\label{fade}
\end{equation}
or, expressed for a generic agent~$n$:
\begin{equation}
\widehat \theta_n(t) {=}  \sum_{m = 1}^N W_{n m}(t) \left( \widehat \theta_m( t {-} 1 ) {+} \frac{1}{t}  c_m \left( y_m(t) {-} \overline y_m(t{-}1) \right) \right).
\label{fadeagent}
\end{equation}
About properties on the matrices $W(t)$ that make FADE succeed, we will say much more in the next section~\ref{subsecwmatrices}.

To conclude, we interpret each iteration of FADE~\eqref{fadeagent} in terms of sensing and communication.  The iteration~\eqref{fadeagent} can be seen as unfolding in two halves: in the first half, each agent $m = 1, \ldots, N$ updates its estimate by absorbing  its measurement, $\overline \theta_m( t  ) = \widehat \theta_m( t - 1 ) + \frac{1}{t}  c_m \left( y_m(t) - \overline y_m(t-1) \right)$, thus using its sensing resource; in the second half, each agent $m = 1, \ldots, N$ sends its updates $\overline \theta_m(t)$ to its neighbors, thus using is communication resource. Upon receiving the updates $\overline \theta_m(t)$ from its neighbors, each agent~$n$ combines  these updates with its own, $\widehat \theta_n(t) = \sum_{m = 1}^N W_{n m}(t) \overline \theta_m(t)$.

\vspace*{2ex}
\noindent\textbf{Comparing the FADE algorithm with the CI algorithm.} 
We now compare the form of the FADE algorithm in~\eqref{fade} with the consensus+innovations (CI) algorithm from~\cite{SoummyaKar_GLU}, holding in mind the optimal estimator in~\eqref{vecform} as a reference point. We will see that, in a certain sense, FADE is closer to the idealized estimator~\eqref{vecform}. Let $\widetilde \theta_m(t)$ be the estimate of the parameter $\theta$ that the CI algorithm produces at agent~$m$ and at time~$t$; letting $\widetilde \theta(t ) = \left( \widetilde \theta_1(t), \ldots, \widetilde \theta_N(t) \right) \in {\mathbf R}^N$ be the vector of estimates across the network, we have (see~\cite{SoummyaKar_GLU})
\begin{equation}
\widetilde \theta(t) {=} \left( I_N {-} \beta(t) L(t) \right) \widetilde \theta(t{-}1) {+} \alpha(t) C \left( y(t) {-} {\mathcal H}  \widetilde \theta(t{-}1) \right).
\label{ciupdate}
\end{equation}
Here,  $L(t)$ is the laplacian matrix of the graph ${\mathcal G}(t) = \left( {\mathcal V}, {\mathcal E}(t) \right)$, that is, a $N\times N$ matrix filled with zeros save for the nondiagonal entries correponding to edges in ${\mathcal E}(t)$ (which are filled with $-1$) and the diagonal entries (which are filled with the number of neighbors of the corresponding agent, thus making $L(t) {\mathbf 1} = {\mathbf 0}$);  the matrix ${\mathcal H} \in {\mathbf R}^{N\times N}$ is diagonal with $n$th diagonal entry equal to $h_n$, and both $\alpha(t)$ and $\beta(t)$ are positive step sizes which obey certain requirements: for our purposes, $\alpha(t) = 1/t$ and $\beta(t) = 1/t^r$ with $0 < r < 1/2$ do. Now, letting $W(t) = I_N - \beta(t) L(t)$ and $\alpha(t) = 1/t$, we can rewrite~\eqref{ciupdate} as
\begin{equation}
\widetilde \theta(t) = W(t) \widetilde \theta(t-1) + \frac{1}{t} C \left( y(t) - H  \widetilde \theta(t-1) \right).
\label{ciupdate2}
\end{equation}
When we compare  both the FADE~\eqref{fade} and the CI~\eqref{ciupdate2} updates with the ideal update~\eqref{vecform}, two main differences spring up: first, even if the communication graph was static and fully connected (so, with laplacian matrix  $L = N I_N - {\mathbf 1} {\mathbf 1}^T$), the CI algorithm would differ from the optimal form~\eqref{vecform}, whereas the FADE update and the optimal one would become the same (for we could assign $W(t) = J$); second, the rightmost term in the optimal recursion~\eqref{vecform}---that is, the \textit{innovation} term $y(t) - \overline y( t-1)$---finds itself  replaced by $y(t) - H \widetilde \theta(t-1)$ in the CI algorithm, whereas it is left intact in FADE. So, the algorithm most faithful to the idealized estimator~\eqref{vecform} is FADE. This gives us an inkling of why FADE estimates $\theta$ with an accuracy closer to the accuracy of a central node outstandingly sooner than the CI algorithm, as the numerical simulations in section~\ref{sc:simulation} attest.

\subsection{FADE algorithm for a vector of parameters}
\label{sec:vector}
Extending the FADE algorithm~\eqref{fade} to a vector of parameters $\theta \in {\mathbf R}^d$ with $d > 1$ is plain. 
Recall the FADE update for a scalar parameter, at each agent~$n$ given in~\eqref{fadeagent}. 

In the case of a vector of parameters, the scalar $h_i \in {\mathbf R}$ becomes a matrix $H_i \in {\mathbf R}^{d_i \times d}$: recall~\eqref{eq:obs}. Accordingly, we can upgrade the scalar $c_m \in {\mathbf R}$ to the matrix \begin{equation} C_m =  \left( (1 / N) \sum_{i = 1}^N H_i^T H_i \right)^{-1} H_m^T \in {\mathbf R}^{d \times d_m} \label{matrixCn} \end{equation} and the FADE update to 
\begin{eqnarray} \lefteqn{\widehat \theta_n(t)  =} \nonumber \\ & & \sum_{m = 1}^N W_{n m}(t) \left( \widehat \theta_m(t{-}1) {+} \frac{1}{t} C_m \left( y_m(t) {-} \overline y_m(t{-}1) \right) \right),  \nonumber \\ \label{unpack2} \end{eqnarray}
where $\widehat \theta_m(t) \in {\mathbf R}^d$. This is the general FADE algorithm at each agent~$n$; or,  repacked in matrix form,
\begin{equation}
\widehat \theta(t) = \left( W(t) \otimes I_d \right) \left( \widehat \theta( t - 1 ) + \frac{1}{t}  C \left( y(t) - \overline y(t-1) \right) \right),
\label{fade2}
\end{equation}
where $\widehat \theta(t) = \left( \widehat \theta_1(t)^T, \ldots, \widehat \theta_N(t)^T \right)^T \in {\mathbf R}^{d N}$, $\otimes$ is Kronecker product, $C  \in {\mathbf R}^{dN \times (d_1 + d_2 + \cdots + d_N)}$ is a block-diagonal matrix with $n$th block equal to $C_n$,
$y(t) = \left( y_1(t)^T, \ldots, y_N(t)^T \right)^T \in {\mathbf R}^{d_1 + \cdots + d_N}$, and $\overline y(t) = \left( \overline y_1(t)^T, \ldots, \overline y_N(t)^T \right)^T \in {\mathbf R}^{d_1 + \cdots + d_N}$.

\subsection{The weight matrices $W(t)$}
\label{subsecwmatrices}

We now give conditions on the weight matrices $W(t)$ that allow FADE to succeed. 

First, recall that each $W(t)$ mirrors the sparsity of the edge-set ${\mathcal E}(t)$. That is, entry $W_{mn}(t)$ is nonzero if and only if there is an edge between agents~$n$ and $m$ in the edge-set ${\mathcal E}(t)$.

We assume that each $W(t)$ is symmetric $\left( W(t) = W(t)^T \right)$; has nonnegative entries ($W_{nm}(t) \geq 0$); and is row-stochastic, i.e., the entries in each of its rows sum to one ($W(t) {\mathbf 1} = {\mathbf 1}$). We also assume that each diagonal entry of $W(t)$ is positive: $W_{nn}(t) > 0$ for all $n$. Note that the consensus matrix $J$ (which $W(t)$ is meant to replace in~\eqref{vecform}) has all these properties. 

\vspace*{2ex}
\noindent\textbf{Metropolis weights.} \label{mw-weights}
A simple way to make sure these properties hold for each matrix $W(t)$ is to choose the entries of $W(t)$ as in the Metropolis rule~\cite{boyd2004_metropolis}: $W_{nm}(t) = 1 / \left( \text{max}\left\{d_n(t), d_m(t) \right\}+1 \right)$, if agents $n$ and $m$ are neighbors in ${\mathcal E}(t)$;  \\ $W_{nm}(t) = 1 - \sum_{m \in {\mathcal N}_n(t)} W_{nm}(t)$,  if  $n = m$; and $W_{nm}(t) = 0$, otherwise.
Here, ${\mathcal N}_n(t) = \left\{ m\,:\, \{ n , m \} \in {\mathcal E}(t) \right\}$ is the neighborhood of agent~$n$ (in the graph ${\mathcal G}(t) = \left( {\mathcal V}, {\mathcal E}(t) \right)$, and $d_n(t)$ is the degree of agent~$n$, i.e., the number of its neighbors (the cardinality of the set ${\mathcal N}_n(t)$). The Metropolis rule allows each agent $n$ to compute its weights ($W_{nm}(t)$ for $m = 1, \ldots, N$) locally: agent~$n$ ignores the global edge-set ${\mathcal E}(t)$, a fit property in practice, for otherwise agent~$n$ would need to know which channels turned out in far away corners of the network at each time step~$t$; in the Metropolis tule, agent~$n$ needs to know only its degree and the degrees of its neighbors (an easy-to-get information that can be passed by the neighbors themselves).

The Metropolis rule, then, associates to each edge-set ${\mathcal E}(t)$ a weight matrix $W(t)$.  Because the edge-set ${\mathcal E}(t)$ is random and takes values in a finite collection of edge-sets $\left\{ {\mathcal E}_1, \ldots, {\mathcal E}_K \right\}$ with probability $\pi_k = {\mathbf P}\left( {\mathcal E}(t) = {\mathcal E}_k \right)$, it follows likewise that $W(t)$ is random and  takes values in a finite collection of weighting matrices, say, $\left\{ W_1, \ldots, W_K \right\}$, with corresponding probability $\pi_k = {\mathbf P}\left( W(t) = W_k \right)$. Note also that, as a consequence,  the sequence $W(t)$, $t = 1, 2, \ldots,$ is i.i.d. In sum, we have the following assumption on the weight matrices.

\begin{assumptionno}[(Weight matrices)]
Each weight matrix $W(t)$ in~\eqref{fade2} mirrors the sparsity of the edge-set ${\mathcal E}(t)$. Also, each $W(t)$ has a positive diagonal and is symmetric, nonnegative, and row-stochastic. Moreover, each $W(t)$ takes values in a finite set $\left\{ W_1, \ldots, W_K \right\}$ with probability $\pi_k={\mathbf P}\left( W(t) = W_k \right)$ (where $\pi_k = {\mathbf P}\left( {\mathcal E}(t) = {\mathcal E}_k \right)$), and the sequence $ W(t)$, $t = 1, 2, \ldots$ is i.i.d.\! . \label{assump4}
\end{assumptionno}

This assumption, together with assumption~\ref{assump2} on average connectivity of the edge-sets ${\mathcal E}(t)$, guarantees key properties for two matrices that will prove important in the theoretical analysis of FADE (see  next section~\ref{tfade}): the \textit{average} weighting matrix $\overline W = {\mathbf E}\left( W(t) \right)=  \sum_{k = 1}^K \pi_k W_k$ and the \textit{average off-consensus} matrix \begin{equation} \widetilde W = {\mathbf E}\left( \widetilde W(t)^T \widetilde W(t) \right) = \sum_{k = 1}^K \pi_k \widetilde W_k^T \widetilde W_k, \label{imp2appa} \end{equation}
with \begin{equation} \widetilde W(t) = ( I_N - J) W(t) (I_N - J ) \label{impforappa} \end{equation} and
$\widetilde W_k = ( I_N - J) W_k (I_N - J )$. 
The key properties are stated in the following lemma.

\begin{lemma} \label{lemma1}
Let assumptions~\ref{assump2} and~\ref{assump4} hold. Then, $\overline W$ is a primitive matrix and $\widetilde W$ is a contraction matrix.
\end{lemma}

Recall that a primitive matrix is a square nonnegative matrix $A$ such that, for some positive integer~$m$, the matrix $A^m$ is  positive (i.e., each entry of $A^m$ is a positive number).
In our case, note that $\overline W$ is a symmetric matrix, which makes all of its eigenvalues real-valued; moreover, because $W {\mathbf 1} = {\mathbf 1}$, one of these eigenvalues is the number $1$ with the vector ${\mathbf 1}$ as its associated eigenvector. Now, given that $\overline W$ is a nonnegative matrix and that lemma~\ref{lemma1} states $\overline W$ is also a primitive matrix, it follows from standard Perron-Frobenius theory (see \cite{horn2012matrix}) that $1$ is, in fact, the dominant eigenvalue:  if $\lambda \neq 1$ is another eigenvalue of $\overline W$, then $| \lambda | < 1$. Finally, the matrix $\widetilde W$ being a contraction means that its spectral radius is strictly less than one, $\rho\left( \widetilde W \right) < 1$.

The proof of lemma~\ref{lemma1} is omitted because these these general properties are well-known and follow from classic Perron-Frobenius theory: for example, \cite{PhDThesis} shows that  $\overline W$ is a primitive matrix in Proposition 1.4, and that $\widetilde W$ is a contraction matrix in p. 35.

\section{Theoretical analysis of FADE}
\label{tfade}

In this section, we state the two chief properties that FADE enjoys: almost sure convergence to the true vector of parameters, and asymptotic unbiasedness and efficiency.

We start by establishing almost sure (a.s.) convergence. Let $\widehat \theta_n(1), \widehat \theta_n(2), \widehat \theta_n(3),  \ldots,$  be the sequence of estimates of the vector of parameters $\theta$ that FADE produces at agent~$n$: see~\eqref{unpack2}. Note that this sequence is random because both the measurements and the edges are random. Theorem~\ref{T1} states that the sequence converges almost surely  to the correct vector of parameters~$\theta$.

\begin{theorem}[FADE converges almost surely] \label{T1} Let assumptions~\ref{assumpnoise},~\ref{assumpgo},~\ref{assump2}, and~\ref{assump4} hold. Then, $\lim_{t \rightarrow \infty} \widehat \theta_n(t) = \theta$, a.s., for $n = 1, 2, \ldots, N$.
\end{theorem}
\begin{IEEEproof}[Proof] See appendix~\ref{app:proofth1}.
\end{IEEEproof}

We now pass to asymptotic unbiasedness and efficiency. Asymptotic unbiasedness means that the sequence of estimates $\widehat \theta_n(1), \widehat \theta_n(2), \widehat \theta_n(3), \ldots,$ becomes unbiased, that is, ${\mathbf E}\left( \widehat \theta_n(t) \right)$ converges to $\theta$, as $t \rightarrow \infty$. Asymptotic efficiency means that the mean-square error (MSE) of each term in the sequence of estimates decays \textit{at the same rate} as the optimal estimator. Specifically, let $\widehat \theta_{\text{ML}}(t)$ be the optimal (maximum-likelihood) estimator---the estimator that runs at a central node---, given by \begin{equation} \widehat \theta_{\text{ML}}(t) = \sum_{n = 1}^N \frac{1}{N} C_n \overline y_n(t), \label{optimalml} \end{equation}
where $C_n$ is defined in~\eqref{matrixCn} and~$\overline y_n(t)$ in~\eqref{avey}. (For a scalar parameter, $\theta \in {\mathbf R}$, this estimator coincides with the one given in~\eqref{optimalestimator}.) With standard tools~\cite{kay_estimation_theory}, it is easy to show that the optimal  estimator is unbiased at all times, ${\mathbf E}\left( \widehat \theta_{\text{ML}}(t) \right) = \theta$, and its MSE, given by $\text{MSE}\left( \widehat \theta_{\text{ML}}(t) \right) 
 = {\mathbf E} \left( \left\| \widehat \theta_{\text{ML}}(t) - \theta \right\|^2 \right)$, decays to zero as follows: \begin{equation}\text{MSE}\left( \widehat \theta_{\text{ML}}(t) \right) = \frac{\text{tr}\left( \left( \sum_{n = 1}^N H_n^T H_n \right)^{-1} \right)}{t}. \label{mseoptimalml} \end{equation}
Asymptotic efficiency of FADE means that the decay rate of the MSE of the FADE estimate $\widehat \theta_m(t)$ at any agent matches the decay rate of MSE of the optimal estimator at the central node. The result is stated precisely in the next theorem.

\begin{theorem}[FADE is asymptotically unbiased and efficient] \label{T2} Let assumptions~\ref{assumpnoise},~\ref{assumpgo},~\ref{assump2}, and~\ref{assump4} hold. Then, $\lim_{t \rightarrow \infty} {\mathbf E} \left( \widehat \theta_n(t) \right) = \theta,$
and \begin{equation} \lim_{t \rightarrow \infty} \frac{\text{MSE}\left( \widehat \theta_n(t) \right)}{\text{MSE}\left( \widehat \theta_{\text{ML}}(t) \right)} = 1, \label{msefade} \end{equation}
for $n = 1, 2, \ldots, N$.
\end{theorem}
\begin{IEEEproof}[Proof] See appendix~\ref{app:proofth2}.
\end{IEEEproof}

In this sense, FADE succeeds in making \textit{all} agents as powerful as the central node.

\section{Numerical simulations} \label{sc:simulation}

We compare three  estimators: the proposed FADE estimator $\widehat \theta(t)$ (given in~\eqref{fade2}),  the state-of-art CI estimator $\widetilde \theta(t)$ from \cite{SoummyaKar_GLU}  (given in~\eqref{ciupdate} for scalar parameters), and the centralized estimator $\widehat \theta_{\text{ML}}(t)$ (given in~\eqref{optimalml}).

We compare the estimators in two kinds of simulations. In the first kind of simulations, section~\ref{firstset}, we look at almost sure convergence (theorem~\ref{T1}); we compare the speed at which the estimators converge to the true vector of parameters $\theta$. In the second kind of simulations, section~\ref{secondset}, we look at MSEs (theorem~\ref{T2}); we compare the speed at which the accuracy of all estimators, as measured by their MSEs, goes to zero.

\vspace*{2ex}

\noindent\textbf{Simulation setup.} To compare the estimators, we set up a dense network and a sparse one. Both  networks consist of a set ${\mathcal V}$ of $N = 50$ agents. 

Also, both networks have their edges changing randomly over time. For the dense network, the random edge-set ${\mathcal E}(t)$ takes values in a finite collection of $K = 15$ edge-sets, $\left\{ {\mathcal E}_1, {\mathcal E}_2, \ldots, {\mathcal E}_K \right\}$, with this collection satisfying assumption~\ref{assump2}; that is, the graph ${\mathcal G} = \left( {\mathcal V}, \bigcup_{k = 1}^K {\mathcal E}_k \right)$, which results from overlaying all  edge-sets in the collection $\left\{ {\mathcal E}_1, {\mathcal E}_2, \ldots, {\mathcal E}_K \right\}$,  is connected. We call this network dense because ${\mathcal G}$ is dense. Specifically, about $79\%$ pairs of agents have an edge between them in the edge-set $\cup_{k = 1}^K {\mathcal E}_k$ and, for each ${\mathcal E}_k$, the average degree (number of neighbors) of an agent  is about six. As for the sparse network, the corresponding graph ${\mathcal G}$ connects directly just  $28\%$ pairs of agents, and the average degree of the agents per ${\mathcal E}_k$ drops to a number close to one. 

Whether the network is dense or sparse, agent~$n$ measures the vector of parameters $\theta = ( 100, 120, 70, 90, 200 ) \in {\mathbf R}^5$ through the same linear-gaussian sensing model $y_n(t) = H_n \theta + v_n(t)$, where  $v_n(t)$ is standard gaussian noise and $H_n \in {\mathbf R}^{8\times 5}$. We made each matrix $H_n$ to have only rank~$4$; this is to make sure that no single agent could identify $\theta$, even if given an infinite supply of measurements. Thus, the vector of parameters $\theta$ is identifiable only through collaboration (the set of matrices $H_n$, $n = 1, \ldots, N$ chosen secures global observability, see assumption~\ref{assumpgo}).
 
Finally, we use the Metropolis weights for FADE (see section~\ref{mw-weights}), and the step-sizes $\alpha(t) = 1/t$ and $\beta(t) = 1/t^{0.05}$ for CI (see~\eqref{ciupdate}).

\subsection{First kind of simulations: almost sure convergence}
\label{firstset}

All three estimators---FADE $\widehat \theta(t)$, CI $\widetilde \theta(t)$, and the centralized one $\widehat \theta_{\text{ML}}(t)$---converge almost surely (a.s.) to~$\theta$ as the number of communications  grows unbounded: 
\begin{equation}
\lim_{t \rightarrow \infty} \widehat \theta_n(t) = \lim_{t \rightarrow \infty} \widetilde \theta_n(t) = \lim_{t \rightarrow \infty} \widehat \theta_{\text{ML}}(t) = \theta, \quad \quad a.s. , \label{tas} 
\end{equation}
for any agent~$n$.
For the optimal estimator, \eqref{tas} follows at once  from~\eqref{optimalml} and the strong law of large numbers (which assures $\lim_{t \rightarrow \infty} \overline y_n(t) = H_n \theta$ almost surely); for FADE, see the proof of theorem~\ref{T1}; for CI, see~\cite{SoummyaKar_GLU}.

But the theoretical analysis fails to  tell us \textit{how fast} the convergence in~\eqref{tas} occurs. The reason is that the analysis is asymptotic; it explains only  what happens in the long run, for $t = \infty$.  In this  remote horizon---after an \textit{infinite} number of communications---\eqref{tas} shows that all estimators look the same. We ignore, however, how the estimators compare in a more realistic horizon: after a number of communications that is \textit{practical}. 

\vspace*{2ex}

\begin{figure}[!t]
\centering
\includegraphics[height=0.5\columnwidth]{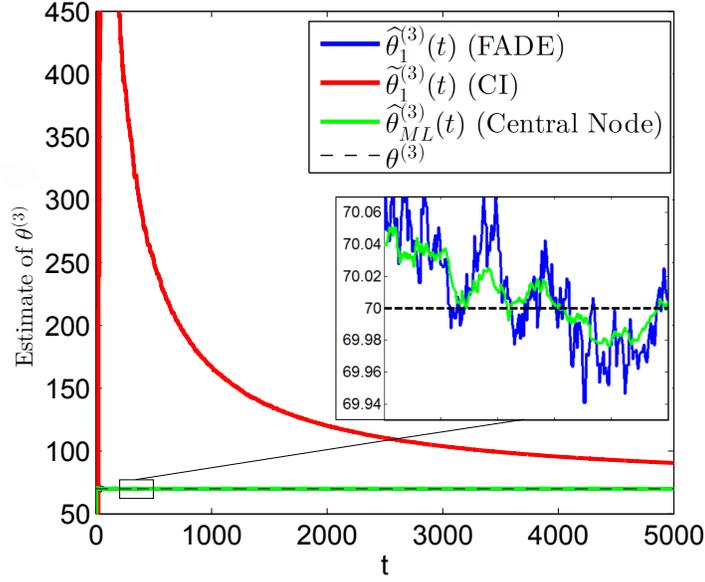}
\caption{How the three estimates at agent~$1$---$\widehat \theta_1^{(3)}(t)$ for FADE, $\widetilde \theta_1^{(3)}(t)$ for CI, and $\widehat \theta_{\text{ML}}^{(3)}(t)$ for the centralized estimator---approach their limit $\theta^{(3)} =70$ in the sparse network, along one path of $5000$ time-steps. The FADE estimate is on par with the quick centralized estimate: the magnified box resolves the time-steps between $250$ and $500$. The CI estimate  has a considerable relative error of about $20\%$, even after $5000$ communications; in contrast, both FADE and the centralized estimator attain a scant relative error of about $0.1\%$, as early as after $250$ steps.}
\label{as_conv}
\end{figure}

\noindent\textbf{Results for the sparse network.} We use the challenging sparse network to find the speeds at which the three estimators approach the limit~\eqref{tas}. Specifically, we focus on agent~$1$ and track its estimate of the third entry of the vector of parameters $\theta = \left( 100, 120, 70, 90, 200 \right)$,  as yielded by the three estimators:
 $\widehat \theta_1^{(3)}(t)$ for FADE, $\widetilde \theta_1^{(3)}(t)$ for CI, and $\widehat \theta_{\text{ML}}^{(3)}(t)$ for the centralized estimator. According to~\eqref{tas}, all estimates go to $\theta^{(3)} = 70$,
\begin{equation}
\lim_{t \rightarrow \infty} \widehat \theta_1^{(3)}(t) = \lim_{t \rightarrow \infty} \widetilde \theta_1^{(3)}(t) = \lim_{t \rightarrow \infty} \widehat \theta_{\text{ML}}^{(3)}(t) = 70, \quad a.s. . 
\end{equation}
Figure~\ref{as_conv} reveals, however, that the estimates go to $\theta^{(3)} = 70$ at stunningly different speeds: CI lags appreciably behind, while  FADE goes hand in hand with the swift centralized estimator.

For this example, we blinded agent~$1$ to the third entry of $\theta$, $\theta^{(3)}$, by filling the third column of the sensing matrix $H_1$  with zeros. This means that $\theta^{(3)}$ has no bearing on the measurements of agent~$1$, and that $\theta^{(3)}$ can only be learned quickly at agent~$1$ through effective teamwork. Figure~\ref{as_conv} shows that FADE delivers such effective coordination, for it supplies  agent~$1$ with an accurate guess of the missing parameter, promptly.

\subsection{Second kind of simulations: scaled MSEs}
\label{secondset}

The MSEs of the three estimators decay to zero  at the rate ${\mathcal O}(1 / t)$.   Specifically,  when we scale the MSEs by the number of time-steps~$t$, the scaled MSEs converge to the \textit{same} limit:
\begin{eqnarray} \lim_{t \rightarrow \infty} t\, \text{MSE}\left( \widehat \theta_n(t) \right) & = & \lim_{t \rightarrow \infty} t\, \text{MSE}\left( \widetilde \theta_n(t) \right) \nonumber \\ & = & \lim_{t \rightarrow \infty} t\, \text{MSE}\left( \widehat \theta_{\text{ML}}(t) \right) \nonumber \\ & = & \text{tr}\left( \left( \sum_{i = 1}^N H_i^T H_i \right)^{-1} \right), \nonumber \\  \label{tmses} \end{eqnarray}
for any agent~$n$.
For the optimal estimator, \eqref{tmses} follows from~\eqref{mseoptimalml}; for FADE, see the proof of~\eqref{msefade} in appendix~\ref{app:proofth2}; for~CI, see~\cite{SoummyaKar_GLU}.

For the optimal estimator, the convergence in~\eqref{tmses} is instantaneous. This is because $t\, \text{MSE}\left( \widehat \theta_\text{ML}(t) \right)$ is constant (so, already equal to its limit from the first time-step); see~\eqref{mseoptimalml}. For FADE and CI, however, the available theoretical analysis is unable to inform about \textit{how fast} the convergence in~\eqref{tmses} takes place.

As the following numerical results show, the proposed FADE estimator converges outstandingly faster.

\vspace*{2ex}

\begin{figure}[!t]
\centering
\includegraphics[height=0.5\columnwidth]{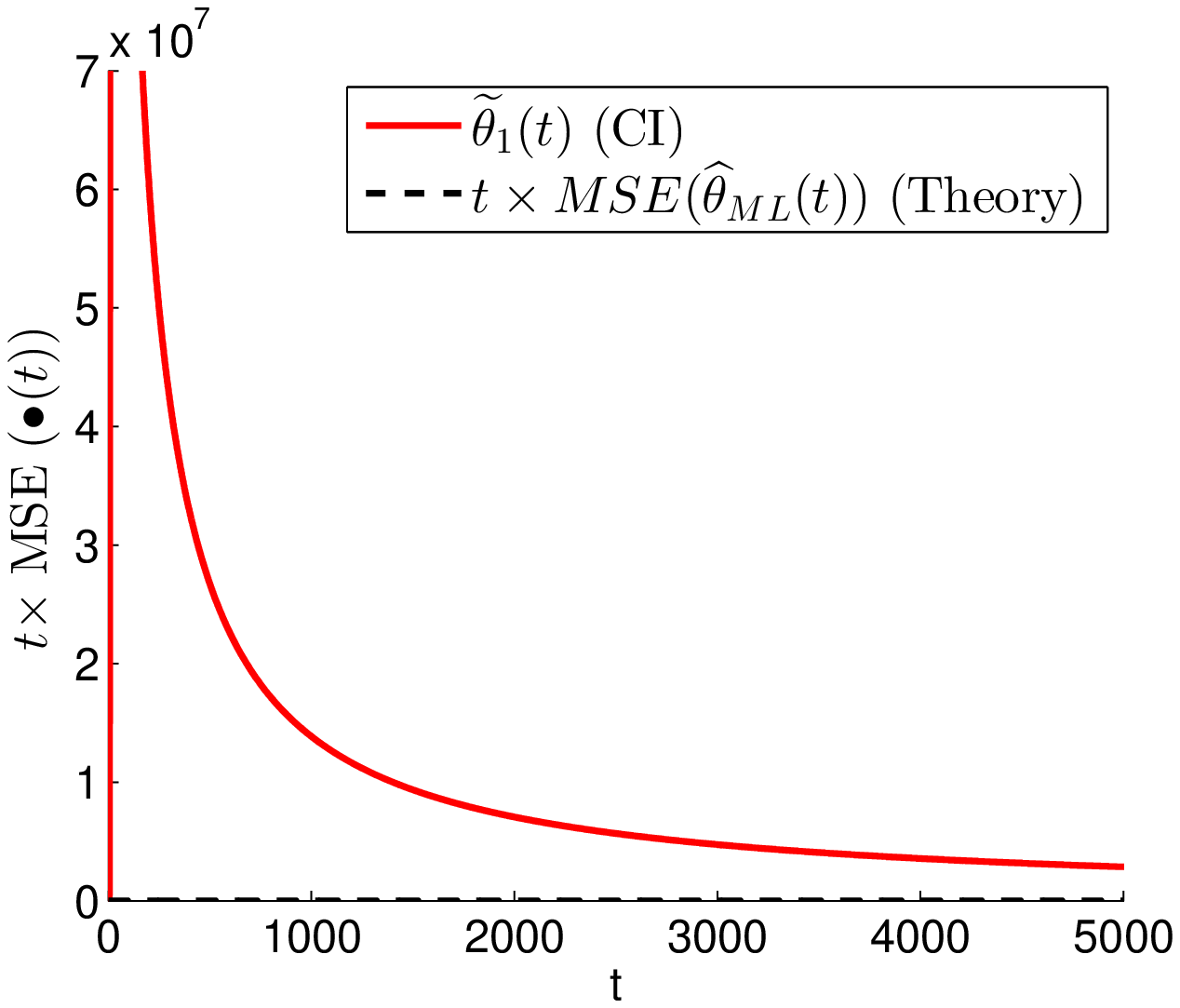}\\
\includegraphics[height=0.5\columnwidth]{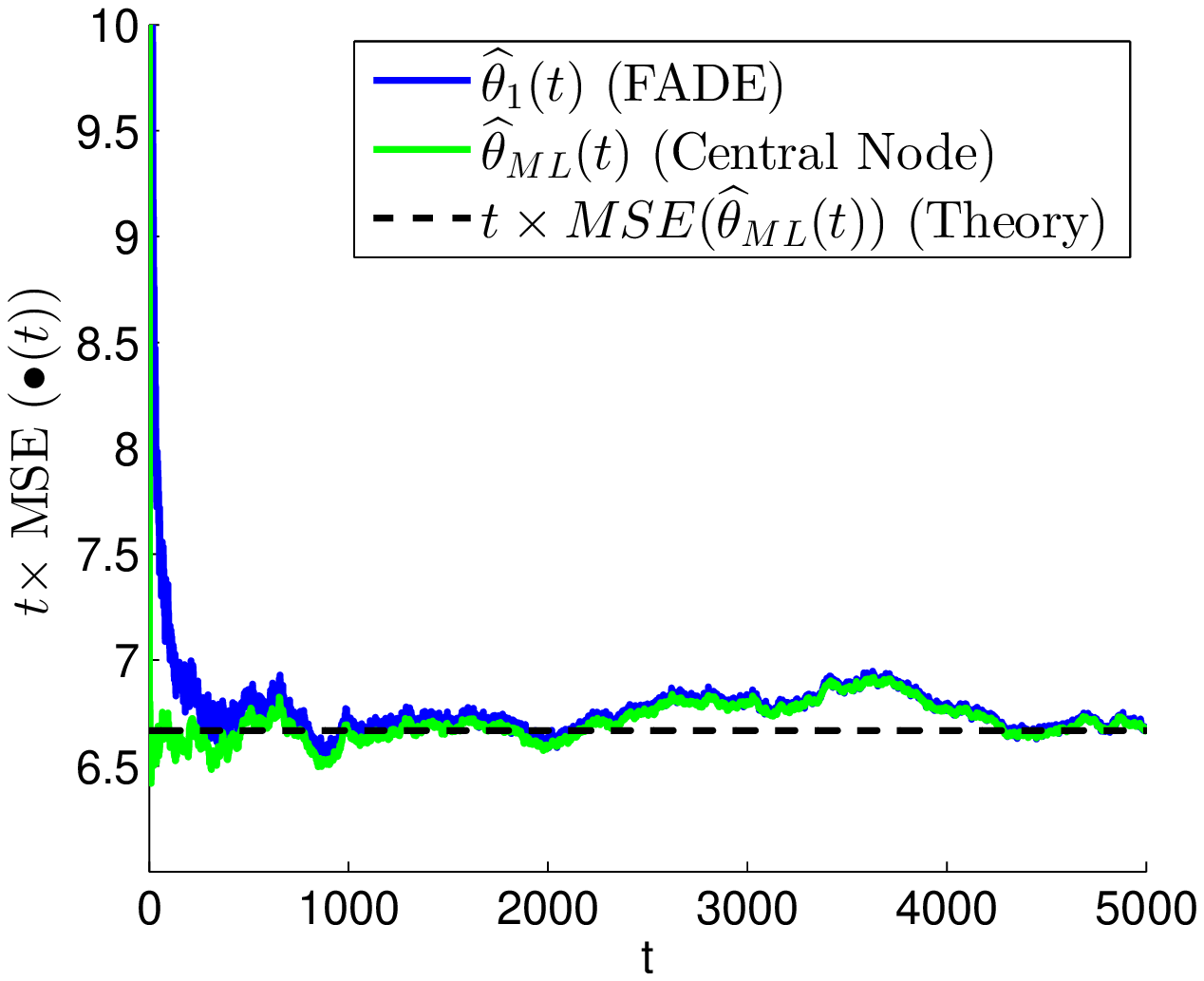}
\caption{This figure compares throughout time, at agent~$1$ in the \textit{dense} network,  the scaled MSE of the optimal centralized estimator with the MSEs of two distributed estimators: the state-of-art  CI estimator and the proposed FADE estimator. The top plot shows the CI and centralized estimators; the bottom plot shows the FADE and centralized estimators. The CI estimator is significantly outdistanced by the proposed FADE estimator, with FADE  keeping up with the optimal centralized estimator: note that, from the top to the bottom plot, the range of the vertical axis shrinks by six orders of magnitude.}
\label{dense_network}
\end{figure}

\noindent\textbf{Results for the dense network.} We look at the scaled MSEs that the three estimators (FADE, CI, and the centralized one) provide at agent~$1$ for the dense network model. We run $1000$ Monte-Carlos, each one consisting of a stretch of $5000$ time-steps, and we average at the end the Monte-Carlos to find out how the scaled estimators---$t\, \text{MSE}\left( \widehat \theta_1(t) \right)$ for FADE, $t\, \text{MSE}\left( \widetilde \theta_1(t) \right)$ for CI, and  $t\, \text{MSE}\left( \widehat \theta_{\text{ML}}(t) \right)$ for the centralized estimator---behave throughout the time-steps~$t$. Figure~\ref{dense_network} shows the results: the proposed FADE estimator follows closely the quick centralized estimator, while the CI estimator is off by six orders of magnitude.

\vspace*{2ex}

\begin{figure}[!ht]
\centering
\includegraphics[height=0.5\columnwidth]{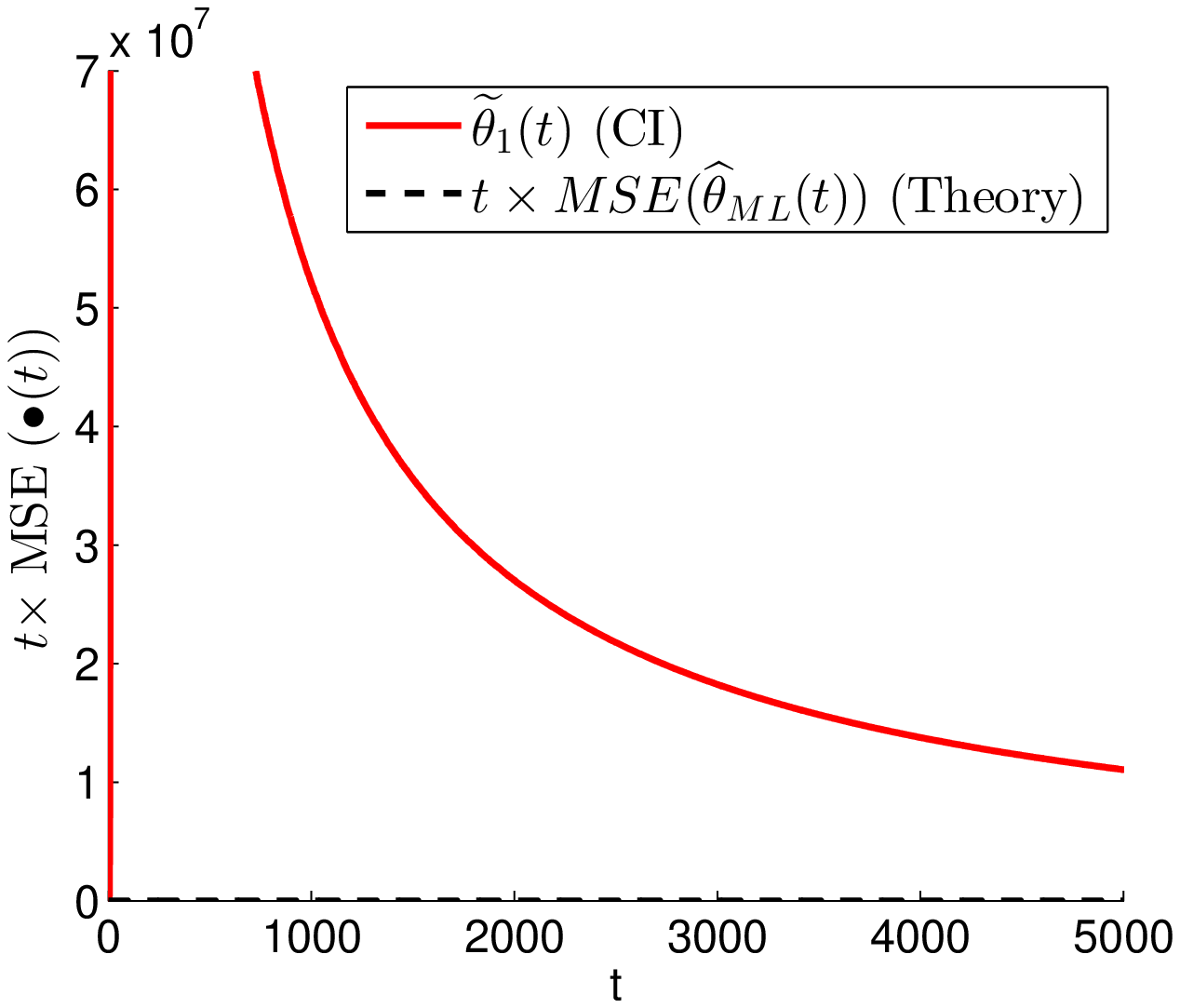}\\
\includegraphics[height=0.5\columnwidth]{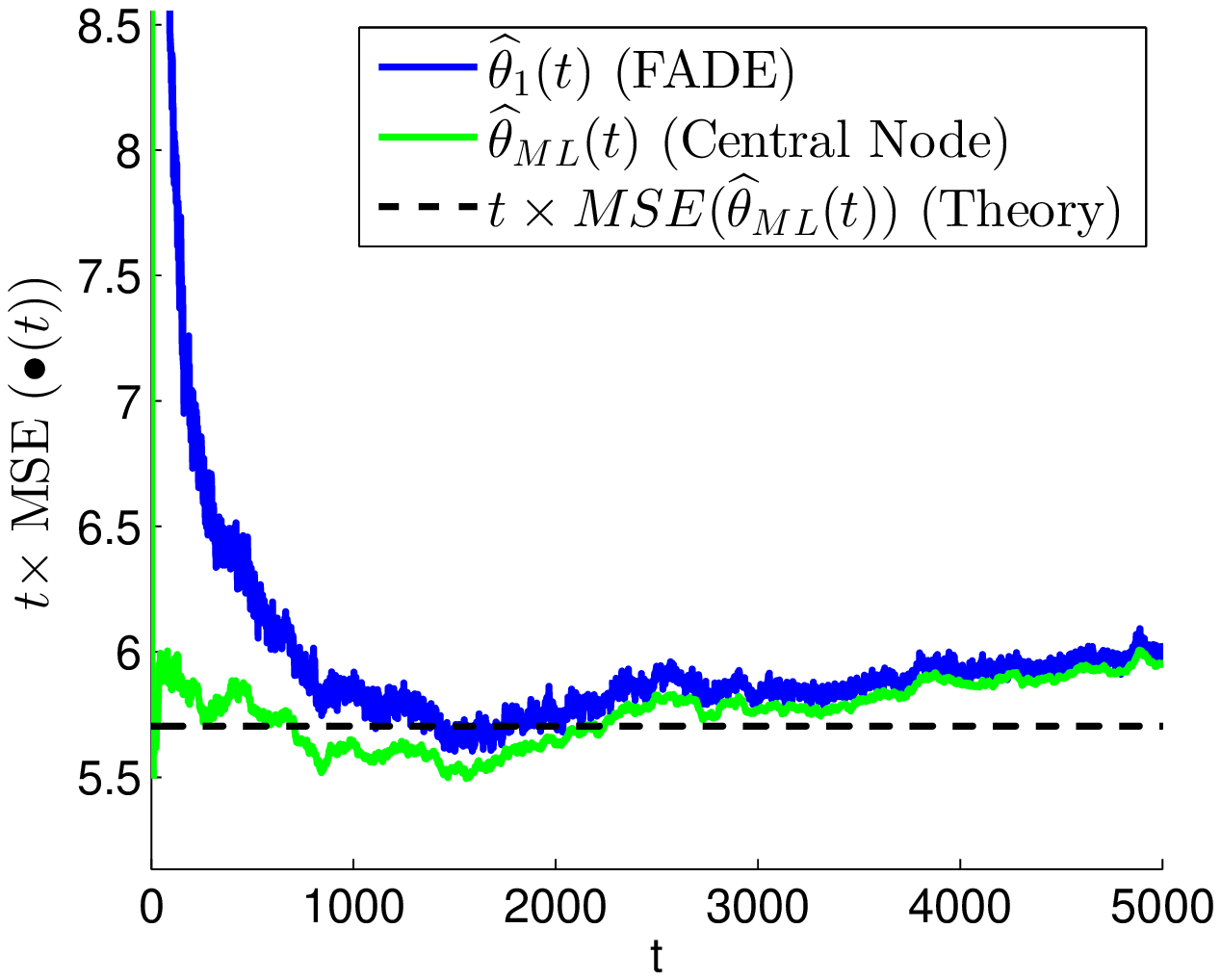}
\caption{Similar to figure~\ref{dense_network} but with a \textit{sparse} network: note that, from the top to the bottom plot, the range of the vertical axis shrinks by seven orders of magnitude.} 
\label{sparse_network}
\end{figure}

\noindent\textbf{Results for the sparse network.} For the sparse network, the difference in performance grows even larger, to seven orders of magnitude; see Figure~\ref{sparse_network}.

\section{Conclusions} \label{sc:conclusion}
We proposed a new algorithm for distributed parameter estimation with linear-gaussian measurements. Our algorithm, called FADE (Fast and Asymptotically efficient Distributed Estimator), is simple to derive and copes with communication networks that change randomly. FADE comes with strong theoretical guarantees: not only is it strongly consistent, but also asymptotically efficient. Compared with a state-of-art consensus+innovations algorithm, FADE yields estimates with significantly smaller mean-square-error (MSE)---in numerical simulations, FADE features estimates with MSEs that can be six or seven orders of magnitude smaller.


\appendices


\section{Proof of theorem~\ref{T1}} \label{app:proofth1}

\noindent\textbf{Scalar parameter.}
We will prove the theorem for the case of a scalar parameter---$\theta \in {\mathbf R}$---for clarity. The proof for the general vector case $\theta \in {\mathbf R}^d$, $d > 1$, is immediate and left to the reader. For the case of a scalar parameter, the network-wide measurement vector $y(t) = \left( y_1(t), y_2(t), \ldots, y_N(t) \right)^T \in {\mathbf R}^N$ is
$y(t) = h \theta + v(t)$,
where $h = ( h_1, h_2, \ldots, h_N )^T  \in {\mathbf R}^N$ is a nonzero vector, thanks to assumption~\ref{assumpgo}.

The FADE algorithm is 
\begin{equation}
\widehat \theta(t) = W(t) \left( \widehat \theta( t - 1 ) + \frac{1}{t}  C \left( y(t) - \overline y(t-1) \right) \right),
\label{fade_appa}
\end{equation}
for $t \geq 1$, with $\widehat \theta(0) = 0$. Also, recall that $C$ is a diagonal matrix with $n$th-entry  $c_n =  \left( (1/N) \sum_{i = 1}^N h_i^2  \right)^{-1} h_n$,
and
$\overline y(t) = \frac{1}{t} \sum_{s = 1}^t y(s)$  with $\overline y(0) = 0$.

The goal is to show that \begin{equation} \lim_{t \rightarrow \infty} \widehat \theta(t) = \theta {\mathbf 1},  \quad a.s. . \label{proofgoal} \end{equation} 

\vspace*{2ex}

\noindent\textbf{The in-consensus and off-consensus orthogonal decomposition.} Any vector $u \in {\mathbf R}^N$ can be decomposed as an orthogonal sum of two vectors: one vector aligned with the vector ${\mathbf 1} \in {\mathbf R}^N$, and the other vector orthogonal to ${\mathbf 1}$. That is,  \begin{equation}
u = u_\top {\mathbf 1} + u_{\bot}
\label{decomp}
\end{equation}
where the scalar $u_\top = {\mathbf 1}^T u / N \in {\mathbf R}$ and the vector $u_\bot = u - u_\top {\mathbf 1}$. It follows that $u_\bot = (I_N - J) v$ (recall that $J = {\mathbf 1} {\mathbf 1}^T / N$) is the consensus matrix. The vector $u_\top {\mathbf 1}$ is called the \textit{in-consensus} component of $u$; the vector $u_\bot$, its \textit{off-consensus} component.

Decomposing as such the vector \begin{equation} \widehat \theta(t) = \widehat \theta_\top(t) {\mathbf 1} + \widehat \theta_\bot(t), \label{decomp} \end{equation} we see that~\eqref{proofgoal} ammounts to 
\begin{equation}
\lim_{t \rightarrow \infty} \widehat \theta_\top(t) = \theta ,  \quad a.s. , \label{goal1}
\end{equation}
and
\begin{equation}
\lim_{t \rightarrow \infty} \widehat \theta_\bot(t) = 0,  \quad a.s. . \label{goal2}
\end{equation}

We will prove~\eqref{goal1} and~\eqref{goal2} separately.

\subsection{Proof of~\eqref{goal1}}

From assumption~\ref{assump4}, each matrix $W(t)$  is symmetric and row-stochastic, which means that each $W(t)$ is also column-stochastic: ${\mathbf 1}^T W(t) = {\mathbf 1}^T$. So, multiplying~\eqref{fade_appa} on the left by ${\mathbf 1}^T / N$ gives
\begin{equation}
\widehat \theta_\top(t)  =  \widehat \theta_\top(t-1) + \frac{h^T}{\left\| h \right\|^2} \frac{1}{t} \left( y(t) - \overline y(t-1) \right),
\label{aux_appa} \end{equation}
for $t \geq 1$, with $\widehat \theta_\top(0) = 0$ and $\overline y(0) = 0$.

For $t {=} 1$, we have $\overline y(t{-}1) {=} \overline y(0) {=} 0$, and~\eqref{aux_appa} implies \begin{eqnarray} \widehat \theta_\top(1) & = & \frac{h^T}{\left\| h \right\|^2} \left( h \theta + v(1) \right) \nonumber \\ & = & \theta + \frac{h^T}{\left\| h \right\|^2} v(1). \label{eq1_a} \end{eqnarray}

For $t \geq 2$, we have $\overline y(t-1) = h \theta + \overline v(t-1)$ and~\eqref{aux_appa} implies
\begin{equation}
\widehat \theta_\top(t)  =  \widehat \theta_\top(t-1) + \frac{h^T}{\left\| h \right\|^2} \frac{1}{t} \left( v(t) - \overline v(t-1) \right). \label{eq2_a}
\end{equation}

Rolling the recursion~\eqref{eq2_a} from~\eqref{eq1_a} yields, for $t \geq 1$  \begin{equation} \widehat \theta_\top(t) = \theta + \frac{h^T}{\left\| h \right\|^2} \overline v(t) \ . \label{key_appa} \end{equation} 

Finally, the strong law of large numbers gives $\lim_{t \rightarrow \infty} \overline v(t) = \lim_{t \rightarrow \infty} \frac{1}{t} \sum_{s = 1}^t v(s) = 0$, a.s., which, when plugged in~\eqref{key_appa}, proves~\eqref{goal1}.

\subsection{Proof of~\eqref{goal2}}

\noindent\textbf{A recursive equality for $\widehat \theta_\bot(t)$.} Recall that, by definition,  $\widehat \theta_\bot(t) = (I_N - J ) \widehat \theta(t)$, where $J = {\mathbf 1} {\mathbf 1}^T / N$. Thus, using~\eqref{fade_appa}, we have
\begin{equation}
\widehat \theta_\bot(t) {=} (I_N {-} J )W(t) \left( \widehat \theta( t {-} 1 ) {+} \frac{1}{t}  C \left( y(t) {-} \overline y(t{-}1) \right) \right).
\label{fade_appa2}
\end{equation}
Now, since each $W(t)$ is row- and column-stochastic (${\mathbf 1}^T W(t) = {\mathbf 1}^T$ and $W(t) {\mathbf 1} = {\mathbf 1}$), it follows that \begin{equation} \left( I_N - J \right) W(t) =  \widetilde W(t) \left( I_N - J \right), \label{imp_appa} \end{equation}
where $\widetilde W(t)$ is defined as (recall~\eqref{impforappa}) $\widetilde W(t) = \left( I_N - J \right) W(t) \left( I_N - J \right)$.

Plugging~\eqref{imp_appa} into~\eqref{fade_appa2} gives the recursion
\begin{equation}
\widehat \theta_\bot(t) = \widetilde W(t) \widehat \theta_\bot( t - 1 ) + \frac{1}{t} \widetilde W(t)  C \left( y(t) - \overline y(t-1) \right).
\label{fade_appa3}
\end{equation}
In obtaining~\eqref{fade_appa3}, we also used the identity $\widetilde W(t) \left( I_N - J \right) = \widetilde W(t)$.

\vspace*{2ex}

\noindent\textbf{A recursive inequality for $\left\| \widehat \theta_\bot(t) \right\|^2$}. Fix some positive number $\epsilon > 0$ (which will be set judiciously soon). Using the fact that $\left\| u + v \right\|^2 \leq \left( 1 + \epsilon \right) \left\| u \right\|^2 + \left( 1 + \frac{1}{\epsilon} \right) \left\| v \right\|^2$ holds for generic vectors $u$ and $v$, we deduce from~\eqref{fade_appa3} the inequality
\begin{align}
\left\| \widehat \theta_\bot(t) \right\|^2 &\leq \left( 1 + \epsilon \right) \left\| \widetilde W(t) \widehat \theta_\bot( t - 1 ) \right\|^2 + \nonumber\\
&+ \left( 1 + \frac{1}{\epsilon} \right) \frac{1}{t^2} \left\| \widetilde W(t)  C \left( y(t) - \overline y(t-1) \right) \right\|^2. \label{pass_appa}
\end{align}

Denote the spectral norm (maximum singular value) of a matrix $A$ by  $\left\| A \right\|$ and recall that $\left\| A u \right\| \leq \left\| A \right\| \left\| u \right\|$ for any matrix $A$ and vector $u$. We derive from~\eqref{pass_appa} that
\begin{align}
\left\| \widehat \theta_\bot(t) \right\|^2 &\leq \left( 1 + \epsilon \right)  \widehat \theta_\bot( t - 1 )^T \widetilde W(t)^T \widetilde W(t) \widehat \theta_\bot(t-1) + \nonumber\\
&+ \left( 1 + \frac{1}{\epsilon} \right) \frac{1}{t^2} \left\| \widetilde W(t)  C \right\|^2 \left\| y(t) - \overline y(t-1)  \right\|^2. \label{fade4_appa}
\end{align}

\vspace*{2ex}
\noindent\textbf{A key inequality for ${\mathbf E}\left( \left\| \widetilde \theta_\bot(t) \right\|^2 \, | \, {\mathcal F}(t-1) \right)$.} Let ${\mathcal F}(1) \subset {\mathcal F}(2) \subset {\mathcal F}(3) \subset \cdots$ be the natural filtration, that is, ${\mathcal F}(t)$ is the sigma-algebra generated by all random objects until time~$t$: ${\mathcal F}(t) = \sigma\left( W(1), W(2), \ldots, W(t), v(1), v(2), \ldots, v(t) \right)$.

Note that the random matrix $\widetilde W(t)$ is independent from ${\mathcal F}(t-1)$. Thus, using this fact and  standard properties of conditional expectation (e.g., see~\cite{chung2001course}), we deduce from~\eqref{fade4_appa} that
\begin{align}
&{\mathbf E} \left( \left\| \widehat \theta_\bot(t) \right\|^2 \, {|} \, {\mathcal F}(t{-}1) \right) {\leq} \left( 1 {+} \epsilon \right)  \widehat \theta_\bot( t {-} 1 )^T  \widetilde W \widehat \theta_\bot(t{-}1) {}\nonumber\\
&+ \left( 1 + \frac{1}{\epsilon} \right) \frac{1}{t^2} P {\mathbf E}\left( \left\| y(t) - \overline y(t-1)  \right\|^2 \, |\, {\mathcal F}(t-1) \right), \label{fade5_appa}
\end{align}
where we defined the number $P =  {\mathbf E}\left( \left\|  \widetilde W(t) C \right\|^2 \right)$,
and the matrix $\widetilde W$ was defined in~\eqref{imp2appa} as $\widetilde W =  {\mathbf E}\left( \widetilde W(t)^T \widetilde W(t) \right)$.

We now further refine the two terms in the right-hand side of~\eqref{fade5_appa}: 

\textbf{(i)} note that $\widetilde W$ is a symmetric matrix, which means that its spectral norm and spectral radius coincide. So,
$\widehat \theta_\bot( t - 1 )^T  \widetilde W \widehat \theta_\bot(t-1) \leq \rho\left( \widetilde W \right) \left\| \widehat \theta_\bot( t - 1) \right\|^2$.
This means
\begin{equation}
\left( 1 {+} \epsilon \right)  \widehat \theta_\bot( t {-} 1 )^T  \widetilde W \widehat \theta_\bot(t{-}1) {\leq} \left(1 {-} r\right) \left\| \widetilde \theta_\bot(t{-}1) \right\|^2, \label{bound1_appa}
\end{equation}
where $r = 1 - \left( 1 + \epsilon \right) \rho\left( \widetilde W \right)$.

Recall from lemma~\ref{lemma1} that $\widetilde W$ is contractive, that is, $\rho\left( \widetilde W \right) < 1$. Thus, we can choose $\epsilon > 0$ small enough so that $r > 0$. In the following, we assume that $\epsilon$ has been so chosen;

\textbf{(ii)} for $t \geq 2$, we have $y(t) {-} \overline y(t{-}1)  =  \left( h \theta {+} v(t) \right) {-} \left( h \theta {+} \overline v(t{-}1) \right) =  v(t) {-} \overline v(t{-}1)$,
which implies
\begin{align}
{\mathbf E}&\left( \left\| y(t) {-} \overline y(t{-}1) \right\|^2 \, | \, {\mathcal F}(t{-}1) \right) {=} \nonumber\\
&={\mathbf E}\left( \left\| v(t) \right\|^2 \right) {-} 2 {\mathbf E}\left( v(t) \right)^T \overline v(t{-}1) {+} \left\| \overline v(t{-}1) \right\|^2 \nonumber \\ &= N {+} \left\| \overline v(t{-}1) \right\|^2. \label{bound2_appa}
\end{align}
(For the last equality we used the fact that each component of the additive noise vector $v(t) = \left( v_1(t), v_2(t), \ldots, v_N(t) \right)$ has zero mean and unit variance.)

Plugging~\eqref{bound1_appa} and~\eqref{bound2_appa} in~\eqref{fade5_appa} gives
\begin{align}
{\mathbf E} &\left( \left\| \widehat \theta_\bot(t) \right\|^2 \, {|} \, {\mathcal F}(t{-}1) \right) {\leq} \left\| \widehat \theta_\bot(t{-}1) \right\|^2 {-} r \left\| \widehat \theta_\bot(t{-}1) \right\|^2 {} \nonumber\\
&{+} \left( 1 {+} \frac{1}{\epsilon} \right) \frac{1}{t^2} P \left( N {+} \left\| \overline v(t{-}1) \right\|^2 \right). \label{fade6_appa}
\end{align}

\noindent\textbf{The Robbins-Siegmund lemma.} The Robbins-Siegmund supermartingale convergence lemma~\cite{Robbins1985} states that if $\left( X(t) \right)_{t \geq 1}$, $\left( Y(t) \right)_{t \geq 1}$ and $\left( Z(t) \right)_{t \geq 1}$ are three sequences of nonnegative random variables, each  adapted to a given filtration $\left( {\mathcal F}(t) \right)_{t \geq 1}$ such that ${\mathbf E}\left( X(t)\,|\,{\mathcal F}(t-1) \right) \leq X(t-1) - Y(t-1) + Z(t-1)$ and $\sum_{t = 1}^\infty Z(t) < \infty$ almost surely, then $X(t)$ converges and $\sum_{t = 1}^\infty Y(t) < \infty$ almost surely.

Defining $X(t) = \left\| \widehat \theta_\bot(t) \right\|^2$, $Y(t) = r \left\| \widehat \theta_\bot(t) \right\|^2$, and $Z(t) = \left( 1 + \frac{1}{\epsilon} \right) \frac{1}{(t+1)^2} P \left( N + \left\| \overline v(t) \right\|^2 \right)$, and recalling~\eqref{fade6_appa}, we see that all conditions of the Robbins-Siegmund lemma are satisfied; in particular, $\sum_{t \geq 1} Z(t) < \infty$ is satisfied almost surely because $\overline v(t)$ is bounded almost surely (in fact, owing to the strong law of large numbers, $\overline v(t)$ converges to zero almost surely) and the deterministic series $\sum_{t \geq 1} 1 / t^2$ converges.

From $\sum_{t \geq 1} Y(t) < \infty$, we then conclude that $\left\| \widehat \theta_\bot(t) \right\|^2$ converges to zero almost surely, thus proving~\eqref{goal2}.

\section{Proof of theorem~\ref{T2}} \label{app:proofth2}

\noindent\textbf{Scalar parameter.}
As in the proof  of theorem~\ref{T1} (in appendix~\ref{app:proofth1}), we will show that theorem~\ref{T2} holds for scalar parameters $\theta \in {\mathbf R}$, for clearness; the proof for vector of parameters $\theta \in {\mathbf R}^d$, $d > 1$, is omitted because it is similar. 

For the case of a scalar parameter, FADE is given by~\eqref{fade_appa}. (In this appendix~\ref{app:proofth2}, we will use several identities proved in appendix~\ref{app:proofth1}.) The goal is to show that \begin{equation}
\lim_{t \rightarrow \infty} {\mathbf E}\left( \widehat \theta(t) \right)  =   \theta {\mathbf 1} \label{goalB1} 
\end{equation}
and
\begin{equation}
\lim_{t \rightarrow \infty} \frac{\text{MSE}\left( \widehat \theta_n(t) \right)}{\text{MSE}\left( \widehat \theta_{\text{ML}}(t) \right)}   =   1, \quad \text{for } n = 1, 2, \ldots, N. \label{goalB2} \end{equation}
We prove~\eqref{goalB1} and~\eqref{goalB2} one at a time.

\subsection{Proof of~\eqref{goalB1}}
\label{proofgoalB1}

From~\eqref{decomp}, we have \begin{equation} {\mathbf E}\left( \widehat \theta(t) \right) = {\mathbf E}\left( \widehat \theta_\top(t) \right) {\mathbf 1} + {\mathbf E}\left( \widehat \theta_\bot(t) \right). \label{edecomp} \end{equation}

Since ${\mathbf E}\left( \overline v(t) \right) = 0$, applying the expectation operator to both sides of~\eqref{key_appa} gives \begin{equation} {\mathbf E}\left( \widehat \theta(t) \right) = \theta. \label{edecomp2} \end{equation}

Since ${\mathbf E}\left( \left\| \overline v(t-1) \right\|^2 \right) = N / (t-1)$, applying the expectation operator to both sides of~\eqref{fade6_appa} gives
\begin{align}
{\mathbf E}&\left( \left\| \widehat \theta_\bot(t) \right\|^2 \right) \leq {\mathbf E} \left( \left\| \widehat \theta_\bot(t-1) \right\|^2 \right)  \nonumber\\
&- r {\mathbf E} \left( \left\| \widehat \theta_\bot(t-1) \right\|^2 \right) + \left( 1 + \frac{1}{\epsilon} \right) \frac{P N}{t^2}  \frac{t}{t-1}.
\label{edecomp3}
\end{align}

Inequality~\eqref{edecomp3} is ripe for the application of the Robbins-Siegmund lemma stated at the end of appendix~\ref{app:proofth1}---specifically, its simpler \textit{deterministic} version, where the sequences $\left( X(t) \right)_{t \geq 1}$, $\left( Y(t) \right)_{t \geq 1}$, and $\left( Z(t) \right)_{t \geq 1}$ are deterministic, thereby implying ${\mathbf E}\left( X(t) \,|\, {\mathcal F}(t-1) \right) = X(t)$. In more detail, we can take here $X(t) = {\mathbf E}\left( \left\| \widehat \theta_\bot(t) \right\|^2 \right)$, $Y(t) = r {\mathbf E}\left( \left\| \widehat \theta_\bot(t) \right\|^2 \right)$, and $Z(t) = \left( 1 + \frac{1}{\epsilon} \right) \frac{P N}{t^2}  \frac{t}{t-1}$ to conclude that $\sum_{t \geq 1} Y(t) < \infty$, which implies \begin{equation} \lim_{t \rightarrow \infty} {\mathbf E}\left( \left\| \widehat \theta_\bot(t) \right\|^2 \right) = 0. \label{aBqf} \end{equation}

Finally, because ${\mathbf E}\left( X \right) \leq {\mathbf E}\left( X^2 \right)^{1/2}$ holds for any nonnegative random variable $X$, we have
${\mathbf E}\left( \left\| \widehat \theta_\bot(t) \right\| \right) \leq {\mathbf E}\left( \left\| \widehat \theta_\bot(t) \right\|^2 \right)^{1/2}$,
which, jointly with~\eqref{aBqf}, implies
\begin{equation}
\lim_{t \rightarrow \infty} {\mathbf E}\left( \widehat \theta_\bot(t) \right) = 0.
\label{tBqf}
\end{equation}

Equations~\eqref{edecomp},~\eqref{edecomp2}, and~\eqref{tBqf} prove~\eqref{goalB1}.

\subsection{Proof of~\eqref{goalB2}}
\label{proofgoalB2}

We must prove~\eqref{goalB2}, which is equivalent to prove
\begin{equation}
\lim_{t \rightarrow \infty} \frac{t \text{MSE}\left( \widehat \theta_n(t) \right)}{t \text{MSE}\left( \widehat \theta_{\text{ML}}(t) \right)}   =   1, \quad \text{for } n = 1, 2, \ldots, N. \nonumber \end{equation} 

Note that~\eqref{mseoptimalml} says $t \text{MSE}\left( \widehat \theta_{\text{ML}}(t) \right) = 1 / \left\| h \right\|^2$ which means that we must prove
\begin{equation}
\lim_{t \rightarrow \infty}  t \left\| h \right\|^2 \text{MSE}\left( \widehat \theta_n(t) \right) = 1, \quad \text{for } n = 1, 2, \ldots, N. \label{goalB23} \end{equation}

Now, from the decomposition~\eqref{edecomp} we have $\widehat \theta_n(t) = \widehat \theta_\top(t) + e_n^T \widehat \theta_\bot(t)$, where $e_n \in {\mathbf R}^N$ is the $n$th vector in the canonical basis (that is, $e_n = \left( 0, \ldots, 0, 1, 0, \ldots, 0 \right)^T$ with the $1$ in the $n$th entry). Thus, 
\begin{align}
\text{MSE}&\left( \widehat \theta_n(t) \right)  = {\mathbf E}\left( \widehat \theta_n(t) - \theta \right)^2 \nonumber \\ & = {\mathbf E}\left( \left( \widehat \theta_\top(t) + e_n^T \widehat \theta_\bot(t) - \theta \right)^2 \right) \nonumber \\ & = {\mathbf E}\left( \left( \widehat \theta_\top(t) {-} \theta \right)^2 \right) {+} {\mathbf E}\left( \left( e_n^T \widehat \theta_\bot(t) \right)^2 \right) {+}\nonumber\\ &+ 2 {\mathbf E} \left( \left( \widehat \theta_\top(t) {-} \theta \right) e_n^T \widehat \theta_\bot(t) \right). \nonumber
\end{align}

We prove~\eqref{goalB23} by showing
\begin{equation}
\lim_{t \rightarrow \infty} t \left\| h \right\|^2 {\mathbf E}\left( \left( \widehat \theta_\top(t) - \theta \right)^2 \right) = 1,
\label{stepA_appB}
\end{equation}
\begin{equation}
\lim_{t \rightarrow \infty} t \left\| h \right\|^2 {\mathbf E}\left( \left( e_n^T \widehat \theta_\bot(t) \right)^2 \right)   = 0,
\label{stepB_appB}
\end{equation}
and
\begin{equation}
\lim_{t \rightarrow \infty} t \left\| h \right\|^2 {\mathbf E} \left( \left( \widehat \theta_\top(t) - \theta \right) e_n^T \widehat \theta_\bot(t) \right)   = 0,
\label{stepC_appB}
\end{equation}
for some fixed $n = 1, 2, \ldots, N$.

\vspace*{2ex}
\noindent \textbf{Proof of~\eqref{stepA_appB}.} From~\eqref{key_appa}, we have \begin{equation} {\mathbf E}\left( \widehat \theta_\top(t) - \theta \right)^2 = \frac{1}{t \left\| h \right\|^2}, \label{usus} \end{equation} which implies~\eqref{stepA_appB}.

\vspace*{2ex}
\noindent \textbf{Proof of~\eqref{stepB_appB}.} Since $\left( e_n^T \widehat \theta_\bot(t) \right)^2 \leq \left\| \widehat \theta_\bot(t) \right\|^2$, it suffices to prove
\begin{equation}
\lim_{t \rightarrow \infty} t {\mathbf E}\left( \left\| \widehat \theta_\bot(t) \right\|^2 \right) = 0.
\label{akeyB}
\end{equation}
Multiply both sides of~\eqref{edecomp3} by~$t$ to get
\begin{equation}
\varphi(t) \leq ( 1 {-} r ) \frac{t}{t{-}1} \varphi(t{-}1) + \left( 1+ \frac{1}{\epsilon} \right) \frac{1}{t} P \left( N {+} \frac{N}{t{-}1} \right), \label{athere}
\end{equation}
where $\varphi(t) = t {\mathbf E}\left( \left\| \widehat \theta_\bot(t) \right\|^2 \right)$.

Because $(1 - r ) t / (t -1)$ converges to $1 - r < 1$ as $t \rightarrow \infty$, there exists a $T$ and a $0 \leq \rho < 1$ such that $( 1 - r )  t / ( t - 1 ) \leq \rho < 1$ holds for all $t \geq T$. So, for $t \geq T$,~\eqref{athere} implies
\begin{equation}
\varphi(t) \leq \rho \varphi(t-1) + \left( 1+ \frac{1}{\epsilon} \right) \frac{1}{t} P \left( N + \frac{N}{t-1} \right).\label{athere2}
\end{equation}
It follows that $\left( \varphi(t) \right)_{t}$ is a bounded sequence and, thus, $\varphi = \limsup_{t \rightarrow \infty} \varphi(t)$ is finite. 
Now, using standard properties of the $\limsup$ (such as subadditivity) we get  $\varphi \leq \rho \varphi$, which, given $0 \leq \rho < 1$ and the fact that $\varphi$ is a finite number, means $\varphi = 0$. This proves~\eqref{akeyB}.

\vspace*{2ex}
\noindent \textbf{Proof of~\eqref{stepC_appB}.}
We must prove
$
\lim_{t \rightarrow \infty} t  \left| {\mathbf E} \left( \left( \widehat \theta_\top(t) - \theta \right) e_n^T \widehat \theta_\bot(t) \right)  \right| = 0$.
Since $\left| {\mathbf E}( X ) \right| \leq {\mathbf E}\left( | X | \right)$ for any random variable~$X$, it suffices to prove that
\begin{equation}
\lim_{t \rightarrow \infty} t   {\mathbf E} \left( \left| \left( \widehat \theta_\top(t) - \theta \right) e_n^T \widehat \theta_\bot(t) \right| \right)   = 0.
\label{keyC_appB}
\end{equation}

Now, note that
\begin{align}
\big| \left( \widehat \theta_\top(t) - \theta \right) e_n^T &\widehat \theta_\bot(t) \big|  \leq \left| \widehat \theta_\top(t) - \theta \right| \, \left\| \widehat \theta_\bot(t) \right\| \nonumber \\ & \leq \frac{\lambda}{2} \left( \widehat \theta_\top(t) - \theta \right)^2 + \frac{1}{2 \lambda} \left\| \widehat \theta_\bot(t) \right\|^2, \nonumber
\end{align}
for any fixed $\lambda > 0$. So,
\begin{eqnarray}
\lefteqn{t {\mathbf E} \big( \big| \left( \widehat \theta_\top(t) - \theta \right) e_n^T \widehat \theta_\bot(t) \big| \big)  \leq \frac{\lambda}{2} t {\mathbf E}\left( \widehat \theta_\top(t) - \theta \right)^2 + } \nonumber \\  & &  \frac{1}{2 \lambda} t {\mathbf E} \left( \left\| \widehat \theta_\bot(t) \right\|^2 \right) = \frac{\lambda}{2} \left\| h \right\|^2 + \frac{1}{2 \lambda} t {\mathbf E} \left( \left\| \widehat \theta_\bot(t) \right\|^2 \right), \nonumber \\
\label{usus2}
\end{eqnarray}
where we used~\eqref{usus} to obtain the last equality.

Applying $\limsup_{t \rightarrow \infty}$ to both sides of~\eqref{usus2}---and recalling~\eqref{akeyB}---we  get
\begin{equation}
\limsup_{t \rightarrow \infty} t {\mathbf E} \left( \left| \left( \widehat \theta_\top(t) - \theta \right) e_n^T \widehat \theta_\bot(t) \right| \right) \leq \frac{\lambda}{2} \left\| h \right\|^2. \label{atatzb}
\end{equation}

Finally, since~\eqref{atatzb} is valid for any positive $\lambda > 0$, we conclude 
$\limsup_{t \rightarrow \infty} t {\mathbf E} \left( \left| \left( \widehat \theta_\top(t) - \theta \right) e_n^T \widehat \theta_\bot(t) \right| \right) = 0$, thereby proving~\eqref{keyC_appB}.

%
%

\ifCLASSOPTIONcaptionsoff
  \newpage
\fi



%

\bibliographystyle{IEEEtran}
\bibliography{Trans_SP_MLE}

\begin{thebibliography}{10}
\providecommand{\url}[1]{#1}
\csname url@samestyle\endcsname
\providecommand{\newblock}{\relax}
\providecommand{\bibinfo}[2]{#2}
\providecommand{\BIBentrySTDinterwordspacing}{\spaceskip=0pt\relax}
\providecommand{\BIBentryALTinterwordstretchfactor}{4}
\providecommand{\BIBentryALTinterwordspacing}{\spaceskip=\fontdimen2\font plus
\BIBentryALTinterwordstretchfactor\fontdimen3\font minus
  \fontdimen4\font\relax}
\providecommand{\BIBforeignlanguage}[2]{{%
\expandafter\ifx\csname l@#1\endcsname\relax
\typeout{** WARNING: IEEEtran.bst: No hyphenation pattern has been}%
\typeout{** loaded for the language `#1'. Using the pattern for}%
\typeout{** the default language instead.}%
\else
\language=\csname l@#1\endcsname
\fi
#2}}
\providecommand{\BIBdecl}{\relax}
\BIBdecl

\bibitem{zhao2004wireless}
F.~Zhao and L.~J. Guibas, \emph{Wireless sensor networks: an information
  processing approach}.\hskip 1em plus 0.5em minus 0.4em\relax Morgan Kaufmann,
  2004.

\bibitem{kar2009distributed}
S.~Kar and J.~M. Moura, ``Distributed consensus algorithms in sensor networks
  with imperfect communication: Link failures and channel noise,'' \emph{IEEE
  Transactions on Signal Processing}, vol.~57, no.~1, pp. 355--369, 2009.

\bibitem{dimakis2010gossip}
A.~G. Dimakis, S.~Kar, J.~M. Moura, M.~G. Rabbat, and A.~Scaglione, ``Gossip
  algorithms for distributed signal processing,'' \emph{Proc. of the IEEE},
  vol.~98, no.~11, pp. 1847--1864, 2010.

\bibitem{SocialLearning_2014}
A.~Lalitha, A.~Sarwate, and T.~Javidi, ``Social learning and distributed
  hypothesis testing,'' in \emph{2014 IEEE International Symposium on
  Information Theory}, June 2014, pp. 551--555.

\bibitem{FastConv_Nedic:2017}
A.~Nedic, A.~Olshevsky, and C.~A. Uribe, ``Fast convergence rates for
  distributed non-bayesian learning,'' \emph{IEEE Transactions on Automatic
  Control}, vol.~PP, no.~99, pp. 1--1, 2017.

\bibitem{LargeDev_JX2012}
D.~Bajovic, D.~Jakovetic, J.~M.~F. Moura, J.~Xavier, and B.~Sinopoli, ``Large
  deviations performance of consensus+innovations distributed detection with
  non-gaussian observations,'' \emph{IEEE Trans. on Signal Processing},
  vol.~60, no.~11, pp. 5987--6002, Nov 2012.

\bibitem{SoummyaKar_GLU}
S.~Kar and J.~M.~F. Moura, ``Convergence rate analysis of distributed gossip
  (linear parameter) estimation: Fundamental limits and tradeoffs,'' \emph{IEEE
  Journal of Selected Topics in Signal Processing}, vol.~5, no.~4, pp.
  674--690, Aug 2011.

\bibitem{SoummyaKar_ExpFam}
------, ``Asymptotically efficient distributed estimation with exponential
  family statistics,'' \emph{IEEE Transactions on Information Theory}, vol.~60,
  no.~8, pp. 4811--4831, Aug 2014.

\bibitem{cattivelli2010diffusion_KF}
F.~S. Cattivelli and A.~H. Sayed, ``Diffusion strategies for distributed kalman
  filtering and smoothing,'' \emph{IEEE Transactions on automatic control},
  vol.~55, no.~9, pp. 2069--2084, 2010.

\bibitem{sayed2013diffusion}
A.~H. Sayed, \emph{Diffusion adaptation over networks}, 2013, vol.~3.

\bibitem{chen2012diffusion}
J.~Chen and A.~H. Sayed, ``Diffusion adaptation strategies for distributed
  optimization and learning over networks,'' \emph{IEEE Trans. on Signal
  Processing}, vol.~60, no.~8, pp. 4289--4305, 2012.

\bibitem{mota2012distributed}
J.~F. Mota, J.~M. Xavier, P.~M. Aguiar, and M.~Puschel, ``Distributed basis
  pursuit,'' \emph{IEEE Transactions on Signal Processing}, vol.~60, no.~4, pp.
  1942--1956, 2012.

\bibitem{jakovetic2014fast}
D.~Jakovetic, J.~Xavier, and J.~M. Moura, ``Fast distributed gradient
  methods,'' \emph{IEEE Transactions on Automatic Control}, vol.~59, no.~5, pp.
  1131--1146, 2014.

\bibitem{shi2015extra}
W.~Shi, Q.~Ling, G.~Wu, and W.~Yin, ``Extra: An exact first-order algorithm for
  decentralized consensus optimization,'' \emph{SIAM Journal on Optimization},
  vol.~25, no.~2, pp. 944--966, 2015.

\bibitem{Weng2014efficient}
Z.~Weng and P.~M. Djuric, ``Efficient estimation of linear parameters from
  correlated node measurements over networks,'' \emph{IEEE Signal Processing
  Letters}, vol.~21, no.~11, pp. 1408--1412, 2014.

\bibitem{lopes2008diffusion}
C.~G. Lopes and A.~H. Sayed, ``Diffusion least-mean squares over adaptive
  networks: Formulation and performance analysis,'' \emph{IEEE Trans. on Signal
  Proc.}, vol.~56, no.~7, pp. 3122--3136, 2008.

\bibitem{cattivelli2010diffusion}
F.~S. Cattivelli and A.~H. Sayed, ``Diffusion lms strategies for distributed
  estimation,'' \emph{IEEE Transactions on Signal Processing}, vol.~58, no.~3,
  pp. 1035--1048, 2010.

\bibitem{kar2013consensus}
S.~Kar and J.~M. Moura, ``Consensus+innovations distributed inference over
  networks: cooperation and sensing in networked systems,'' \emph{IEEE Sig.
  Proc. Mag.}, vol.~30, no.~3, pp. 99--109, 2013.

\bibitem{kar2014distributed}
S.~Kar, G.~Hug, J.~Mohammadi, and J.~M. Moura, ``Distributed state estimation
  and energy management in smart grids: A consensus + innovations approach,''
  \emph{IEEE Journal of Selected Topics in Signal Processing}, vol.~8, no.~6,
  pp. 1022--1038, 2014.

\bibitem{kay_estimation_theory}
S.~M. Kay, ``{Fundamentals of statistical signal processing, volume I:
  estimation theory},'' 1993.

\bibitem{boyd2004_metropolis}
S.~Boyd, P.~Diaconis, and L.~Xiao, ``Fastest mixing markov chain on a graph,''
  \emph{SIAM review}, vol.~46, no.~4, pp. 667--689, 2004.

\bibitem{horn2012matrix}
R.~A. Horn and C.~R. Johnson, \emph{Matrix analysis}.\hskip 1em plus 0.5em
  minus 0.4em\relax Cambridge university press, 2012.

\bibitem{PhDThesis}
F.~Iutzeler, ``{Distributed Estimation and Optimization for Asynchronous
  Networks},'' Theses, {Telecom ParisTech}, Dec. 2013.

\bibitem{chung2001course}
K.~L. Chung, \emph{A course in probability theory}.\hskip 1em plus 0.5em minus
  0.4em\relax Acad. press, 2001.

\bibitem{Robbins1985}
H.~Robbins and D.~Siegmund, \emph{"A Convergence Theorem for Non Negative
  Almost Supermartingales and Some Applications"}.\hskip 1em plus 0.5em minus
  0.4em\relax New York, NY: Springer New York, 1985, pp. 111--135.

\end{thebibliography}

%
%

%








\end{document}